\documentclass[11pt,draftclsnofoot,onecolumn]{IEEEtran}

\pdfoutput=1

\usepackage{amsmath,amsfonts,amssymb}
\usepackage{algorithm,algorithmic}
\usepackage{listings,enumerate, graphicx, epsfig, cite, color}
\usepackage{multirow}

\newcommand{\delx}[0]{\partial x }
\newcommand{\dell}[0]{\partial \lambda}

\IEEEoverridecommandlockouts
\title{Dynamic Updating for $\ell_1$ Minimization}
\author{
\IEEEauthorblockN{M. Salman Asif and Justin Romberg} \\
\IEEEauthorblockA{School of Electrical and Computer Engineering \\
Georgia Institute of Technology, Atlanta, Georgia, 30332, USA \\
Email: \{sasif, jrom\}@ece.gatech.edu} \\
\thanks{Manuscript submitted to IEEE Journal of Selected Topics in Signal Processing, March 2, 2009; revised June 26, 2009.}
}

\begin{document}
\maketitle

\begin{abstract}
The theory of compressive sensing (CS) has shown us that under certain conditions, a sparse signal can be recovered from a small number of linear incoherent measurements.
An effective class of reconstruction algorithms involve solving a convex optimization program that balances the $\ell_1$ norm of the solution against a data fidelity term.
Tremendous progress has been made in recent years on algorithms
for solving these $\ell_1$ minimization programs. These algorithms, however, are for the most part static: they focus on finding the solution for a fixed set of measurements. In this paper, we will present a suite of {\em dynamic algorithms} for solving
$\ell_1$ minimization programs for streaming sets of measurements.
We consider cases where the underlying signal changes slightly between measurements, and where new measurements of a fixed signal are sequentially added to the system.
We develop algorithms to quickly update the solution of several different types of $\ell_1$ optimization problems whenever these changes occur, thus avoiding having to solve a new optimization problem from scratch.
Our proposed schemes are based on homotopy continuation,
which breaks down the solution update in
a systematic and efficient way into a small number of linear steps.
Each step consists of a low-rank update and a small number of
matrix-vector multiplications -- very much like recursive least squares.
Our investigation also includes dynamic updating schemes for $\ell_1$ decoding problems, where an arbitrary signal is to be recovered from redundant coded measurements which have been corrupted by sparse errors.
%
%
%
%
\end{abstract}


\begin{IEEEkeywords}
Homotopy, sparse signal recovery, recursive filtering, compressive sensing, $\ell_1$ norm minimization, $\ell_1$ decoding, LASSO, Dantzig selector.
\end{IEEEkeywords}

%
\IEEEpeerreviewmaketitle

\section{Introduction}\label{sec:introduction}
Recovering a signal from a set of linear measurements is a fundamental problem
in signal processing.  We are given measurements $y\in\mathbb{R}^m$ of the form
\begin{equation}\label{eq:y=Ax+e}
y = Ax  + e,
\end{equation}
where $A$ is an $m\times n$ matrix and $e$ is a noise vector.  From these, we wish to reconstruct the unknown signal $x\in\mathbb{R}^n$. The classical solution to this problem is to estimate $x$ from $y$ using least-squares.  Given $y$, we solve
\begin{equation}\label{eq:LS_m>n}
\underset{}{\text{minimize}}\;\;\|A\tilde{x} - y\|_2^2,
\end{equation}
or when $A$ is ill-conditioned
\begin{equation}\label{eq:LS_Tikhonov}
\underset{}{\text{minimize}}\;\;\frac{1}{2}\|A\tilde x - y\|^2_2 + \tau\|\tilde x\|^2_2,
\end{equation}
where $\tau>0$ is a regularization parameter. Each of these minimizers can be found by solving a system of linear equations.
We can interpret the solution to \eqref{eq:LS_Tikhonov} as the estimate which, depending on the value of $\tau$, strikes a balance between the {\em data fidelity} (we want the energy in the mismatch between the simulated measurements $A\tilde x$ of our estimate and the true measurements $y$ to be small) and the {\em complexity} of the estimate (among all estimates with the same measurements, we want the one with minimal energy).

Recent developments in the theory of compressive sensing (CS) have shown us that under certain conditions, dramatic gains can be had by promoting sparsity
instead of minimizing energy.  There are two classes of problems:

\noindent{\bf CS:} In this case, the matrix $A$ is underdetermined, and the
signal $x$ is sparse.  To promote sparsity in the solution, we penalize the
$\ell_1$ norm of the estimate, solving
\begin{equation}\label{eq:Lasso}
\underset{}{\text{minimize}}\;\;\frac{1}{2}\|A\tilde x-y\|^2_2 + \tau\|\tilde
x\|_1.
\end{equation}
For certain types of measurement matrices (namely, matrices that obey a
type of uncertainty principle \cite{CandesTao_2006_NearOptimal}), \eqref{eq:Lasso} comes with a number of performance guarantees \cite{Candes_2006_StableRecovery, Tropp_2006_JustRelax, Donoho_2006_StableRecoveryOvercomplete, Donoho_2006_CS, CandesPlan_2008_NearIdealModelSelection, Zhu_2008_StableRecoveryIT}.
In particular, if $x$ is sparse enough and there is no noise, \eqref{eq:Lasso} will recover $x$ exactly as $\tau\rightarrow 0$ even though $A$ is underdetermined; the recovery can also be made stable when the measurements are made in the presence of noise with an appropriate choice of $\tau$.  There are also several variations on \eqref{eq:Lasso} which use slightly different penalties for the
measurement error.  We will also be interested in one of these variations, the Dantzig Selector \cite{CandesTao_2007_DS} given in \eqref{eq:DS} below.

\noindent{\bf Decoding:} In this case, the matrix $A$ is overdetermined, and
the error $e$ is sparse.  To account for this, we solve
\begin{equation}\label{eq:L1_decode}
\underset{}{\text{minimize}}\;\;\|A\tilde x-y\|_1
\end{equation}
in place of \eqref{eq:LS_m>n}.  There are again a number of performance
guarantees for \eqref{eq:L1_decode} that relate the number of errors we can correct (number of non-zero entries in $e$) to the number of measurements we have collected (rows in $A$)  \cite{CandesTao_2005_DecodingLP,rudelson05ge}.
If the matrix consists of independent Gaussian random variables, then the
number of errors we can correct (and hence recover $x$ exactly) scales with the
amount of oversampling $m-n$.

These $\ell_1$ minimization programs are tractable, but solving them is more
involved than least-squares.  In this paper, we will be interested in how
solutions to these problems change as 1) the signal we are measuring changes by
a small amount, and 2) new measurements of the signal are added.  We will
present a suite of algorithms that avoid solving these programs from scratch each
time we are given a new set of measurements, and instead quickly update the
solution.
We will constrain our discussion to small and medium scale problems, where the matrices are stored explicitly and linear systems of equations are solved exactly (within machine precision) using direct methods.
We begin with a brief review of how updating works in the least-squares
scenario.

\subsection{Update of least-squares}\label{subsec:L2_problems}


When the $m\times n$ matrix $A$ has full column rank (is overdetermined), the least squares problem
\eqref{eq:LS_m>n} has a unique solution $\hat x_0$ found by solving a system of linear equations:
\begin{equation}\label{eq:xhat_LS}
	\hat{x}_0 = (A^TA)^{-1}A^Ty.
\end{equation}
There is a variety of ways to compute $\hat x_0$, including iterative methods that have the potential to return an approximate solution at relatively low cost, but in general the computational cost involved for an exact solution is $O(mn^2)$. Typical direct methods for solving \eqref{eq:LS_m>n} involve Cholesky or QR decompositions \cite{Golub_1996_MatrixComputation,Bjorck_1996_NumericalLS_book}.
If we have already computed the QR factorization for $A$ (or Cholesky factorization for $A^TA$), then there is not much marginal cost in recovering additional signals measured with the same matrix $A$.  We can simply use the already computed factorization for the new set of measurements at a cost of $O(mn)$.

There is also an efficient way to update the solution if we add (or remove) a small number of measurements to the system.  Assume that we have solved \eqref{eq:LS_m>n} using \eqref{eq:xhat_LS} to get the estimate $\hat x_0$ with our current set of $m$ measurements $y$.  Now suppose that we get one new measurement given as $w= bx+d$, where $b$ is a row vector and $d\in\mathbb{R}$ denotes noise in the new observation. The system of equations becomes
\begin{equation}\label{eq:y=Ax+e_w=bx+d}
\begin{bmatrix} y\\w \end{bmatrix} = \begin{bmatrix} A\\b \end{bmatrix} x +
\begin{bmatrix} e\\d \end{bmatrix},
\end{equation}
and the least-squares solution $\hat x_1$ obeys: $(A^TA+b^Tb)\hat{x}_1 = (A^T y + b^T w)$. A naive way to compute $\hat{x}_1$ would be to solve this new system of equations from scratch. But we can avoid this computationally expensive task by using rank-1 updates, reducing the cost of computing the new solution from $O(mn^2)$ to $O(mn)$.

The new solution $\hat{x}_1$ can be written in terms of the previous solution $\hat{x}_0$ using the matrix inversion lemma (also known as the Sherman-Woodbury-Morrison formula)
\footnote{In practice, we will want to update the Cholesky or QR factorizations, rather than the explicit inverse of $A^TA$, as the Sherman-Woodbury-Morrison formula can become numerically unstable if the matrices are not well-conditioned.  Here we discuss the update in terms of the explicit inverses to simplify the exposition, and to separate the main concept --- that the solution of a system of equations can be efficiently updated --- from its implementation.  For a detailed discussion of methods for low-rank updates, see \cite{Bjorck_1996_NumericalLS_book}.}.
The matrix $(A^TA + b^Tb)^{-1}$ can be computed from $(A^TA)^{-1}$ in the following manner.  With $P_0 =
(A^TA)^{-1}$, we set
\begin{equation}
\label{eq:infomat} P_1 := (A^TA + b^Tb)^{-1} = P_0 - K_1bP_0,
\quad\text{where}\quad K_1 = P_0b^T(1+bP_0b^T)^{-1},
\end{equation}
and the new estimate can be written as
\begin{equation}\label{eq:solupdate}
\hat{x}_1 = \hat{x}_0 + K_1(w-b\hat{x}_0).
\end{equation}
Note that $1+bP_0b^T$ in \eqref{eq:infomat} is a scalar, and so the essential cost of the updating procedure is a few matrix-vector multiplies.
Thus given the new measurement $w$, we can find the new solution in $O(mn)$
computations.
%

The goal of this paper is to develop a similar methodology for updating the
solutions to a suite of $\ell_1$ minimization programs.  This will not be as
straightforward as in the least-squares case, but we will see that we can move
between solutions using a series of low rank updates similar to
\eqref{eq:infomat}, \eqref{eq:solupdate}.

\subsection{$\ell_1$ problems}
\label{subsec:L1_problems}

In this section, we give a brief overview of the four types of $\ell_1$ minimization programs for which we will develop dynamic updating algorithms.

A large body of literature has arisen around the problem of reconstructing a
sparse signal from a limited number of measurements.
The essence of this theory, which goes under the name of {\em compressive
sensing}, is that if the $m\times n$ matrix $A$ is {\em incoherent}, then we can reliably
estimate $x$ about as well as if we observed its $m/\log n$ most significant
components directly.  The technical conditions for this incoherence
property basically state that $A$ has to be close to an isometry when it
operates on sparse signals \cite{CandesTao_2006_NearOptimal}.
There are several manners in which these types of
matrices can be generated, the easiest of which is to simply draw the entries
of $A$ independently from a concentrated (e.g.\ Gaussian) distribution \cite{CandesTao_2006_NearOptimal,Donoho_2006_CS}.

We will discuss two optimization programs for sparse signal recovery.  The
first is \eqref{eq:Lasso}, which goes by the name of basis pursuit denoising (BPDN) \cite{Chen_99_BasisPursuit} in signal processing and is the lagrangian formulation of the LASSO \cite{Tibshirani_1996_LASSO}, a well-known tool for model selection in statistics.
Solving \eqref{eq:Lasso} is robust in that it is stable both in the presence of noise and to the fact that the signal may not be exactly sparse
\cite{Candes_2006_StableRecovery,Tropp_2006_JustRelax,Donoho_2006_StableRecoveryOvercomplete}.
Methods for computing the solution to BPDN can be found in
\cite{Chen_99_BasisPursuit,KimBoyd_2007_L1_ls,Figueiredo_2007_GPSR,hale2008fpc,Yin_2008_Bregman,Efron_2004_LARS,OsbornePresnell_2000_NewApproachLasso}.

Related to, but subtly different than, BPDN is the Dantzig selector (DS)
\cite{CandesTao_2007_DS}.  Instead of requiring that the residual $A\tilde x-y$
for a candidate estimate $\tilde x$ have small energy, it asks instead that the residual should not be too correlated with any of the columns of $A$.  Given the measurements $y$, the DS solves
\begin{equation}\label{eq:DS}
\text{minimize} \;\; \|\tilde x\|_1 \quad \text{subject to} \quad \|A^T(A\tilde
x-y)\|_\infty \le \tau,
\end{equation}
for some relaxation parameter $\tau > 0$.  For incoherent $A$, the DS guarantees a
near-optimal estimate of a sparse signal when the measurements are made in the
presence of Gaussian noise.  Algorithms for solving \eqref{eq:DS} can be found
in \cite{l1magic,James_2007_DASSO,Asif_2008_MSThesis}.

While we can compress a sparse signal by applying an underdetermined incoherent
matrix, we can also protect a general signal against sparse errors by applying
an overdetermined incoherent matrix.  If we take $m=Cn$ incoherent measurements of an arbitrary signal $x$, where $C> 1$, and add a sparse error $e$ that has fewer
than $\rho(C)\cdot m$ non-zero terms, where $\rho(C)$ is a constant that
depends on $C$, then solving the optimization program \eqref{eq:L1_decode} will
recover $x$ exactly \cite{CandesTao_2005_DecodingLP,rudelson05ge}.  This result depends only on the number of nonzero terms
in $e$, and not on their magnitude. Another way to interpret the action of $A$
is as a channel encoding which can correct a certain number of (arbitrarily large) errors.

This recovery can also be made robust to small errors present throughout all of
the measurements \cite{CandesRandall_2008_RobustEC}.  Suppose that we measure
\begin{equation}\label{eq:y=Ax+e_RobustDecode}
y = Ax + e + q_y,
\end{equation}
where $A$ is the $m\times n$ coding matrix with $m > n$, $e$ is a sparse error
vector (the gross errors), and $q_y$ is a non-sparse error vector whose
entries are relatively small.  To account for both types of error, we solve
\begin{align}\label{eq:Robust_EC}
\underset{}{\text{minimize}}\; \; &\tau \|\tilde
e \|_1 + \frac{1}{2}\|\tilde q_y\|_2^2 \quad \text{subject to} \quad A\tilde x
+ \tilde e + \tilde q_y = y,
\end{align}
which can be rewritten as
\begin{equation}\label{eq:Robust_EC_Lasso}
\text{minimize} \; \; \tau \|\tilde e \|_1 + \frac{1}{2} \|Q(\tilde e -
y)\|_2^2,
\end{equation}
where $Q$ is a matrix whose rows span the
null space of $A^T$, $QA = 0$; one particular choice is $Q := I - A(A^TA)^{-1}A^T$. This problem is similar to BPDN, and its solution gives us an estimate $\hat e$ of the error. %
The decoded message can then be found using $\hat x = (A^TA)^{-1}A^T(y-\hat e)$.

\subsection{Overview}

The goal of this paper is to develop {\em dynamic algorithms} for solving these
types of $\ell_1$ minimization programs.  We will characterize how their
solutions change when a small number of new measurements are added, and (in the case of the BPDN and DS) when the signal changes.
In doing this, we will see that moving from one solution to the next can be broken down as a series of linear problems which can in turn be solved with a series of low-rank updates.
Our approach is based on \emph{homotopy continuation principle}, which
we describe in Section~\ref{sec:Homotopy}.  The main idea of the homotopy
framework is to slowly change from one optimization program to another by
varying a (carefully placed) parameter in such a manner that we can trace the
solution path. In Section~\ref{sec:DynamicX}, we see how to apply this
principle to update the solution to the BPDN and DS as the signal we are measuring
changes.  In Sections~\ref{sec:Dynamic_seq} we see how to update the solutions to the BPDN and DS when a new measurement is added (the former has been independently addressed previously in \cite{Garrigues_2008_NIPS}). Sections~\ref{sec:L1Decoding} and \ref{sec:RobustL1Decoding} turn to the decoding problem, where we see that there are gains to be had by adding measurements in clusters. Section~\ref{sec:Numerical_Examples} contains numerical experiments that demonstrate the effectiveness of these algorithms, and compares dynamic updating to
state-of-the-art $\ell_1$ minimization algorithms which have been ``warm
started''.  MATLAB code for all of the algorithms presented in this paper, along with scripts that reproduce the figures, is publicly available \cite{AR_l1_homotopy_webpage}.

\section{Homotopy}\label{sec:Homotopy}

Homotopy gives us a continuous transformation from one optimization program to another.  The solutions to this string of programs lie along a continuous parameterized path.  The idea is that while the optimization programs may be difficult to solve by themselves, we can trace this path of solutions as we slowly vary the parameter.

A common use for homotopy is to trace the path of solutions as the relaxation parameter changes.  In this section, we give a brief overview of these methods for BPDN and the DS, as many of the ideas are used in our updating algorithms.

%

%
\subsection{Basis pursuit denoising homotopy}
There is an extensively studied \cite{OsbornePresnell_2000_NewApproachLasso,Efron_2004_LARS,Malioutov_2005_HomotopyContinuation} homotopy algorithm associated with the BPDN that traces the solution to \eqref{eq:Lasso} as the parameter $\tau$ changes.  The path is followed by ensuring that certain optimality conditions are being maintained.  To be a solution to \eqref{eq:Lasso}, a vector $x^*$ must obey the following condition \cite{Bertsekas_1999_NonlinearProg_book, Fuchs_2004_OnSparseRep}:
\begin{equation}\label{eq:opt_lasso}
\|A^T(Ax^*-y)\|_\infty \le \tau. \tag{L}
\end{equation}
We can view (L) as a set of $n$ different constraints, one on each entry of the vector of residual correlations $A^T(Ax^*-y)$.
In addition, a sufficient condition for the optimality of $x^*$ is that the
set of locations for which the constraints in \eqref{eq:opt_lasso} are active (i.e. equal to $\tau$) will be the same as the support of $x^*$ (the set of locations for which $x^*$ is non-zero) \cite{Fuchs_2004_OnSparseRep}.  Denoting this set by $\Gamma$, we can write the optimality conditions for any given value of $\tau$ as
\begin{enumerate}[L1.]
\item $A^T_{\Gamma}(A x^* -y) = -\tau z$
\item $\|A^T_{\Gamma^c}(Ax^* -y) \|_\infty < \tau$,
\end{enumerate}
where $A_\Gamma$ is the $m\times |\Gamma|$ matrix formed from the columns of $A$ indexed by $\Gamma$, and $z$ is a $|\Gamma|$-vector containing the signs of $x^*$ on $\Gamma$.  From this we see that $x^*$ can be calculated directly from the support $\Gamma$ and signs $z$ using
\begin{equation*}
x^* = \begin{cases} (A^T_\Gamma A_\Gamma)^{-1}(A_\Gamma^T y - \tau z) &
\text{on }
\Gamma \\
0 & \text{otherwise}.
\end{cases}
\end{equation*}
Thus we can interpret the solution to BPDN as a type of soft-thresholding: given the support $\Gamma$, we first project $y$ onto the range of $A_\Gamma$ and then we subtract $\tau (A_\Gamma^TA_\Gamma)^{-1}z$.  As we change $\tau$, the solution moves along a line with direction $(A_\Gamma^TA_\Gamma)^{-1}z$ until one of two things happens: an element of $x^*$ is shrunk to zero, removing it from the support of $x^*$, or another constraint in \eqref{eq:opt_lasso} becomes active, adding a new element to the support of $x^*$.  At these so-called {\em critical points}, both the support of $x^*$ and the direction of the solution path change.  Also, at any point on the solution path it is straightforward to calculate how much we need to vary $\tau$ to take us to a critical point in either direction.

With these facts in hand, we can solve \eqref{eq:Lasso} by starting with a very large value of $\tau$ (i.e., $\tau > \|A^Ty\|_\infty$), where the solution is the zero vector, and reduce it to the desired value while hopping from one critical point to the next.  At each critical point along this path, a single element is either being added to or removed from $\Gamma$, and the new direction can be computed from the old using a rank-1 update.  Thus multiple solutions over a range of $\tau$ can be calculated at very little marginal cost.

\subsection{Dantzig selector homotopy}
The homotopy algorithm for the Dantzig selector (DS) is similar in principle to the BPDN homotopy \cite{Asif_2008_MSThesis, James_2007_DASSO}. The essential difference in the case of our DS homotopy algorithm is that we have to keep track of both the primal and dual solution for \eqref{eq:DS} as we change $\tau$. The dual problem to the DS in \eqref{eq:DS} can be written as
\begin{align} \label{eq:DS_dual}
&\underset{}{\text{maximize}} \;\; -(\tau \|
\lambda\|_{1}+\langle\lambda,A^Ty\rangle) \quad  \textrm{subject to} \quad
\|A^TA\lambda\|_{\infty} \leq 1 ,
\end{align}
where $\lambda\in \mathbb{R}^n$ is the dual optimization variable.  We can derive the required optimality conditions by recognizing that at the solution, the objectives in \eqref{eq:DS} and \eqref{eq:DS_dual} will be equal, due to strong duality \cite{Boyd_book_ConvexOptimization}.  This fact, along with the complementary slackness property, means that a primal-dual solution pair $(x^*, \lambda^*)$ to \eqref{eq:DS} and \eqref{eq:DS_dual} for any given value of $\tau$ must satisfy the following optimality conditions \cite{Asif_2008_MSThesis}:
\begin{enumerate}[DS1.]
\item $A^T_{\Gamma_\lambda}(A x^* -y) = \tau z_\lambda$
\item $A^T_{\Gamma_x}A \lambda^* = -z_x$
\item $\|A^T_{\Gamma_\lambda^c}(A x^* -y) \|_\infty < \tau$
\item $\|A^T_{\Gamma^c_x} A \lambda^*\|_\infty < 1$,
\end{enumerate}
where $\Gamma_x$ and $\Gamma_\lambda$ are the supports of $x^*$ and $\lambda^*$
respectively, $z_x$ and $z_\lambda$ are the sign sequences of $x^*$ and
$\lambda^*$ on their respective supports. We will call (DS1,DS3) the {\em primal constraints}, and (DS2,DS4) the {\em dual constraints}. From these optimality conditions we can see that the primal and dual solutions can be calculated directly using the supports and sign sequences $(\Gamma_x, \Gamma_\lambda, z_x, z_\lambda)$.
Also we can see that the active primal constraints correspond to the support of dual variable and the active dual constraints correspond to the support of primal variable.

With these facts we can develop the homotopy algorithm for DS in a similar way; we start from a large value of $\tau$ (i.e., $\tau > \|A^Ty\|_\infty$, where the solution is the zero vector) and reduce $\tau$ gradually by updating the support and sign sequence at every critical point. As we change $\tau$, the solution moves along a line in the direction $-(A^T_{\Gamma_\lambda}A_{\Gamma_x})^{-1} z_\lambda$ until one of the two things happens at a new critical point: an element in $x^*$ shrinks to zero (removing an element from the support of $x^*$) or an inactive primal constraint becomes active (adding an element to the support of $\lambda^*$). We call this first phase the \emph{primal update}. This gives us the value of $x^*$ at the new critical point but the value of $\lambda^*$ is still unknown. So we use the information about the change in the support from the primal update phase to find the new value for the dual solution $\lambda^*$ at this critical point, during which either an existing element in $\lambda^*$ shrinks to zero (removing an element from the support of $\lambda^*$) or an inactive dual constraint becomes active (adding an element to the support of $x^*$). We call this second phase the \emph{dual update}. For further details on the DS homotopy see \cite{Asif_2008_MSThesis}.

The homotopy algorithms we discuss below are in many ways similar to the standard BPDN and DS homotopy. In each of them we will introduce a homotopy parameter into the optimization program that gradually incorporates the new measurements as we vary it from $0$ to $1$.  The path the solution takes will again be piecewise linear, and we will jump from critical point to critical point, determining the direction to move using modified version of the optimality conditions L1-L2 and DS1-DS4 above.  Each step will be very efficient, requiring only a few matrix-vector multiplications.
We start with the problem of recovering a time varying sparse signal.

\section{Dynamic update of time varying sparse signal}\label{sec:DynamicX}

In this section we will discuss the problem of estimating a time-varying sparse
signal from a {\em series} of linear measurement vectors.  We expect that the
signal changes only slightly between measurements, so the reconstructions
will be closely related. There are many scenarios where this type of problem
could arise.  For example, in real-time magnetic resonance imaging we want to
reconstruct a series of closely related frames from samples in the frequency
domain \cite{lustig2007sma}.  Another application is channel equalization in
communications, where we are continuously trying to estimate a time varying
(and often times sparse) channel response
\cite{CotterRao_2002_SparseChannelEst}.

Assume that we have solved the BPDN problem \eqref{eq:Lasso} for the system in \eqref{eq:y=Ax+e} for a given value of $\tau$. Now say that the underlying signal $x$ changes to $\breve x$ and we get a new set of $m$ measurements given as
\begin{equation}\label{eq:DynamicX_ymod}
\breve y = A\breve x + \breve e.
\end{equation}
We are interested in solving the following updated BPDN problem
\begin{equation}\label{eq:DynamicX_Lasso}
\text{minimize} \; \; \tau \|\tilde x\|_1 + \frac{1}{2}\|A\tilde x - \breve
y\|_2^2,
\end{equation}
for the same value of $\tau$. Since we expect that the signal changes only slightly between the measurements, the reconstruction will be closely related. Our goal is to avoid solving \eqref{eq:DynamicX_Lasso} from scratch, instead using the information from the solution of \eqref{eq:Lasso} to quickly compute the solution for \eqref{eq:DynamicX_Lasso}. Similarly we are interested in quickly computing the solution of the following updated DS problem
\begin{equation}\label{eq:DynamicX_DS}
\text{minimize}\;\; \|\tilde x\|_1 \quad \text{subject to } \quad
\|A^T(A\tilde x-\breve y)\|_\infty \le \tau,
\end{equation}
by using the information from the solution of \eqref{eq:DS}.

We will develop the homotopy algorithms for updating the solution for
\eqref{eq:DynamicX_Lasso} and \eqref{eq:DynamicX_DS} following three steps. First, we provide a homotopy formulation for the problem moving from one set of measurements to next. Second, we derive the optimality conditions that the solution must obey for each value of the homotopy parameter. Finally, we use these optimality conditions to trace the path towards the new solution.

\subsection{Basis pursuit denoising update}
Let us first look at the dynamic update of the solution for the BPDN problem. Our proposed homotopy formulation is as follows:
\begin{equation}\label{eq:DynamicX_homotopy}
\text{minimize} \; \; \tau \|\tilde x\|_1 + \frac{1-\epsilon}{2}\|A\tilde x -
y\|_2^2+\frac{\epsilon}{2}\|A\tilde x - \breve y\|_2^2,
\end{equation}
where $\epsilon$ is the homotopy parameter. As we increase $\epsilon$ from $0$ to $1$ we move from the solution of the old optimization program \eqref{eq:Lasso} to the solution of the new one \eqref{eq:DynamicX_Lasso}.

By adapting the optimality conditions L1 and L2 from
Section~\ref{sec:Homotopy}, we see that for $x^*$ to be a solution to
\eqref{eq:DynamicX_homotopy} at a given values of $\epsilon$ we must have
\begin{equation}\label{eq:DynamicX_homotopy_constr}
\|A^T(Ax^* -(1-\epsilon)y-\epsilon \breve y) \|_\infty \le \tau,
\end{equation}
or more precisely,
\begin{subequations}
\begin{gather}
A^T_{\Gamma}(A x^* -(1-\epsilon)y-\epsilon \breve y) = -\tau z \tag{\ref{eq:DynamicX_homotopy_constr}a} \label{eq:DynamicX_homotopy_constr1}\\
\|A^T_{\Gamma^c}(Ax^* -(1-\epsilon)y-\epsilon \breve y) \|_\infty < \tau,
\tag{\ref{eq:DynamicX_homotopy_constr}b} \label{eq:DynamicX_homotopy_constr2} 
\end{gather}
\end{subequations}
where $\Gamma$ is the support of $x^*$ and $z$ is its sign sequence on
$\Gamma$. We can see from \eqref{eq:DynamicX_homotopy_constr1} that again the
solution to \eqref{eq:DynamicX_homotopy} follows a piecewise linear path as
$\epsilon$ varies; the critical points in this path occur when an element is
either added or removed from the solution $x^*$.

Suppose that we are at a solution $x_k$ (with support $\Gamma$ and signs $z$) to
\eqref{eq:DynamicX_homotopy} at some critical value of $\epsilon = \epsilon_k$
between zero and one.  To find the direction to move, we will examine how the
optimality conditions behave as $\epsilon$ increases by an infinitesimal amount
from $\epsilon_k$ to $\epsilon_k^+$.  The solution $x_k^+$ at $\epsilon =
\epsilon_k^+$ must obey
\begin{equation}\label{eq:DynamicX_homotopy_constr1p}
A^T_{\Gamma}(A x_{k}^+ -(1-\epsilon_k^+)y-\epsilon_k^+ \breve y) = -\tau z. 
\end{equation}
Subtracting \eqref{eq:DynamicX_homotopy_constr1} from
\eqref{eq:DynamicX_homotopy_constr1p}, the difference between the solutions
$\widetilde \delx = x^+_{k}-x_k$ will be
\begin{equation*}
\widetilde{ \partial x} = \begin{cases} \Delta\epsilon\cdot (A^T_{\Gamma}A_\Gamma)^{-1} A^T_\Gamma(\breve y- y) & \text{on } \Gamma \\
 0 & \text{otherwise,} \end{cases}
\end{equation*}
where $\Delta\epsilon = \epsilon_k^+-\epsilon_k$.  So as $\epsilon$ increases
from $\epsilon_k$, the direction the solution moves is given by
\begin{equation}
\label{eq:DynamicX_delx}
\partial x = \begin{cases} (A^T_{\Gamma}A_\Gamma)^{-1}A^T_\Gamma (\breve y- y) & \text{on } \Gamma \\
 0 & \text{otherwise.} \end{cases}
\end{equation}

With the direction to move given by \eqref{eq:DynamicX_delx}, we need to find
the step-size $\theta$ that will take us to the next critical value of
$\epsilon$.  We increase $\epsilon$ from $\epsilon_k$, moving the solution away
from $x_k$ in the direction $\delx$, until one of the two things happens: one
of the entries in the solution shrinks to zero or one of the constraints in
\eqref{eq:DynamicX_homotopy_constr2} becomes active (equal to $\tau$).  The
smallest amount we can move $\epsilon$ so that the former is true is simply
\begin{equation}
\label{eq:dynx_theta1} \theta^- = \underset{j\in \Gamma}{\text{min}} \left(
\frac{-x_k(j)}{\delx  (j)} \right)_+,
\end{equation}
where $\text{min}(\cdot)_+$ denotes that the minimum is taken over positive
arguments only.  For the latter, set
\begin{subequations}\label{eq:DynamicX_pk_dk}
\begin{gather}
p_k = A^T(A x_k-y +\epsilon_k (y-\breve y)) \\
d_k = A^T (A\delx +y -\breve y).
\end{gather}
\end{subequations}
We are now looking for the smallest stepsize $\Delta\epsilon$ so that $p_k(j)
+\Delta\epsilon\cdot d_k(j) = \pm\tau$ for some $j\in\Gamma^c$.  This is given
by
\begin{equation}
\label{eq:dynx_theta2} \theta^+ = \underset{j\in \Gamma^c}{\text{min}} \left(
\frac{\tau-p_k(j)}{d_k(j)}, \frac{\tau+p_k(j)}{-d_k(j)}  \right)_+.
\end{equation}
So the stepsize to the next critical point is
\begin{equation}
\label{eq:dynx_mintheta} \theta = \min(\theta^+,\theta^-).
\end{equation}

With the direction $\delx$ and stepsize $\theta$ chosen, the next critical
value of $\epsilon$ and the solution at that point will be
\[
\epsilon_{k+1} = \epsilon_k + \theta, \quad  x_{k+1} = x_k + \theta\delx.
\]
The support for new solution $x_{k+1}$ differs from $\Gamma$ by one element.
Let $\gamma^-$ be the index for the minimizer in \eqref{eq:dynx_theta1} and
$\gamma^+$ be the index for the minimizer in \eqref{eq:dynx_theta2}.  If we
chose $\theta^-$ in \eqref{eq:dynx_mintheta}, then we remove $\gamma^-$ from
the support $\Gamma$ and the sign sequence $z$.  If we chose $\theta^+$ in
\eqref{eq:dynx_mintheta}, then we add $\gamma^+$ to the support, and add the
corresponding sign to $z$.

This procedure is repeated until $\epsilon=1$.  A precise outline of the
algorithm is given in Algorithm~\ref{alg:DynamicX} in
Appendix~\ref{app:Pseudo-codes}.

The main computational cost at every homotopy step comes from solving
a $|\Gamma|\times|\Gamma|$ system of equations to compute the direction in
\eqref{eq:DynamicX_delx}, and two matrix-vector multiplications to compute the
$d_k$ for the stepsize.  Since the support changes by a single element from
step to step, the update direction can be computed using a rank-1
update, as described in Section~\ref{subsec:L2_problems}.  As such, the
computational cost of each step is $O(mn)$. 
\subsection{Dantzig selector update}
The homotopy algorithm for dynamic update of DS with time varying signals is very similar to the BPDN update, with the additional requirement of updating both the primal and dual solutions at every homotopy step. Our proposed homotopy formulation is as follows:
\begin{equation}\label{eq:DynamicX_DS_homotopy}
\text{minimize} \;\;\|\tilde x\|_1 \quad \text{subject to } \quad
\|A^T(A\tilde x-(1-\epsilon)y-\epsilon \breve y)\|_\infty \le \tau,
\end{equation}
where $\epsilon$ is the homotopy parameter. The optimality conditions for any primal-dual solution pair $(x^*, \lambda^*)$ to \eqref{eq:DynamicX_DS_homotopy} at a given value of $\epsilon$ can be written as
\begin{subequations}\label{eq:DynamicX_DS_opt}
\begin{gather}
A^T_{\Gamma_\lambda}(A x^* - (1-\epsilon)y-\epsilon\breve y) =
\tau z_\lambda \label{eq:DynamicX_DS_opt1}\\
A^T_{\Gamma_x}A \lambda^* = -z_x \label{eq:DynamicX_DS_opt2}\\
\|A^T_{\Gamma_\lambda^c}(A x^*- (1-\epsilon)y-\epsilon \breve y) \|_\infty < \tau \label{eq:DynamicX_DS_opt3}\\
\|A^T_{\Gamma_x^c} A \lambda_k\|_\infty < 1.\label{eq:DynamicX_DS_opt4}
\end{gather}
\end{subequations}
It can be seen from \eqref{eq:DynamicX_DS_opt1} that the solution $x^*$ to \eqref{eq:DynamicX_DS_homotopy} follows a piecewise linear path w.r.t. $\epsilon$, and there will be some critical points along the homotopy path where the support of $x^*$ and/or $\lambda^*$ change.

\subsubsection{Primal update} Suppose that we are at some critical value of $\epsilon = \epsilon_k$, with primal-dual solution $(x_k, \lambda_k)$ with support and sign sequence $(\Gamma_x, \Gamma_\lambda, z_x, z_\lambda)$. As we change $\epsilon$ from $\epsilon_k$ to $\epsilon_{k}^+$, the solution changes to $x_k^+ = x_k + \Delta \epsilon \delx$, where $\delx$ is given as
\begin{equation}\label{eq:DynamicX_DS_delx}
{ \partial x} = \begin{cases} (A^T_{\Gamma_\lambda}A_{\Gamma_x})^{-1} A^T_{\Gamma_\lambda}(\breve y-y) & \text{on } \Gamma_x \\
 0 & \text{otherwise}, \end{cases}
\end{equation}
and $\Delta \epsilon = \epsilon_k^+-\epsilon_k$. If we start to move in the direction $\delx$ by increasing $\epsilon$ from $\epsilon_k$, at some point either a primal constraint will be activated in \eqref{eq:DynamicX_DS_opt3} (indicating addition of a new element to the support of $\lambda$) or an element in $x_k$ will shrink to zero. We select the smallest step size $\theta$, as described in \eqref{eq:dynx_theta1}, \eqref{eq:dynx_theta2} and \eqref{eq:dynx_mintheta}, such that one of these two things happens. The new critical value of $\epsilon$ will be $\epsilon_{k+1}= \epsilon_k+\theta$ and the new primal solution will be $x_{k+1} = x_k + \theta \delx$.

\subsubsection{Dual update} As we mentioned in the case of standard DS homotopy, we do not yet have the dual solution at this new critical value of $\epsilon$. In the dual update we use the information about the support change from the primal update to find the update direction $\dell$ for the dual vector and consequently the dual solution $\lambda_{k+1}$ at $\epsilon = \epsilon_{k+1}$. Assume that during primal update, a new element entered\footnote{If instead an element was removed from support of $x_k$, we can pick an ``artificial'' index $\gamma \in \Gamma_\lambda$ and treat it as the new element in the support of $\lambda$ with appropriate sign $z_\gamma$.} the support of $\lambda$ at index $\gamma$ with sign $z_\gamma$. Then using \eqref{eq:DynamicX_DS_opt2} we can write the update direction as
\begin{equation*}
\partial \lambda = \begin{cases} -z_\gamma
(A_{\Gamma_x}^T A_{\Gamma_\lambda})^{-1} A_{\Gamma_x}^T a_\gamma & \text{on }
\Gamma_\lambda  \\  z_\gamma & \text{on } \gamma \\  0 & \text{otherwise},
\end{cases}
\end{equation*}
where $a_\gamma$ is the $\gamma${th} column of $A$, $z_\gamma$ is the sign of
$\gamma${th} primal active constraint. This direction ensures that the
dual constraints remain active on $\Gamma_x$ and the sign of new non-zero
element in $\lambda_k^+ = \lambda_k + \delta \dell$ at index $\gamma$ is
$z_\gamma$. As we move our solution $\lambda_k$ in this direction $\dell$ by
increasing the step size $\delta$ from 0, one of two things will happen,
either a nonzero element from $\lambda_k$ will shrink to zero or a dual
constraint in \eqref{eq:DynamicX_DS_opt4} will become active (indicating addition
of a new element in $\Gamma_x$). The smallest step size such that an entry in
$\lambda_k$ shrinks to zero is simply
\begin{equation}
\label{eq:DynamicX_DS_delta1} \delta^- = \underset{j\in \Gamma_\lambda}{\text{min}}
\left( \frac{-\lambda_k(j)}{\dell  (j)} \right)_+.
\end{equation}
The smallest step size such that a constraint in \eqref{eq:DynamicX_DS_opt4}
becomes active is given by
\begin{equation}
\label{eq:DynamicX_DS_delta2} \delta^+ = \underset{j\in \Gamma_x^c}{\text{min}}
\left( \frac{1-a_k(j)}{b_k(j)}, \frac{1+a_k(j)}{-b_k(j)}  \right)_+,
\end{equation}
where $a_k = A^TA \lambda_k$ and $b_k = A^T A\dell$.
The stepsize for the update of dual solution is $\delta = \min(\delta^+,\delta^-)$.
The new dual solution will be $\lambda_{k+1} = \lambda_k + \delta \dell$. The primal and dual support is updated accordingly.

This procedure of primal and dual update is repeated until $\epsilon=1$. 

\section{Dynamic update with sequential measurements} \label{sec:Dynamic_seq}
In this section we will discuss the homotopy algorithms to update the solutions for BPDN and DS as new measurements are added to the system sequentially.
Assume that we have solved the BPDN \eqref{eq:Lasso} for the system in \eqref{eq:y=Ax+e} for some given value of $\tau$. Then we introduce one new
measurement\footnote{We can just as easily remove a measurement by taking $\epsilon$ from 1 to 0 in \eqref{eq:Lasso_updated_homotopy} and \eqref{eq:DS_updated_homotopy}.} $w = bx + d$ as described in \eqref{eq:y=Ax+e_w=bx+d}. We now want to solve the following updated problem
\begin{equation}\label{eq:Lasso_updated}
\text{minimize}\;\;\tau \|\tilde x\|_1 +\frac{1}{2}(\|A\tilde
x-y\|^2_2+|b\tilde x - w|^2),
\end{equation}
for the same value of $\tau$. Similarly for the DS, we want to solve the following updated problem
\begin{equation}\label{eq:DS_updated}
{\text{minimize}} \;\; \|\tilde x\|_1 \;\; \text{subject to} \;\; \|A^T(A\tilde
x-y)+ b^T(b \tilde x-w)\|_\infty \le \tau,
\end{equation}
using the information from the solution of \eqref{eq:DS}.

We will use the same three steps discussed in Section~\ref{sec:DynamicX} to update the solution; first strategically introducing a homotopy parameter, then writing down the appropriate optimality conditions, and finally using the optimality conditions to trace a path to the new solution.

\subsection{Basis pursuit denoising update}
Let us first discuss the homotopy algorithm for the dynamic update of sequential measurements. We note that a similar version of this algorithm has appeared recently in \cite{Garrigues_2008_NIPS}; we include discussion here as it fits nicely into our overall framework, and is closely related to the updating algorithms for the time-varying problem in Section~\ref{sec:DynamicX} and the robust decoding problem in Section~\ref{sec:RobustL1Decoding}.

We incorporate the new measurement gradually by introducing the parameter $\epsilon$, in the homotopy formulation as:
\begin{equation}\label{eq:Lasso_updated_homotopy}
\text{minimize}\;\;\tau \|\tilde x\|_1 +\frac{1}{2}(\|A\tilde
x-y\|^2_2+\epsilon|b\tilde x - w|^2).
\end{equation}
Again, as $\epsilon$ increases from 0 to 1,
we will go from the old problem \eqref{eq:Lasso} to the new one \eqref{eq:Lasso_updated}.

The optimality conditions L1 and L2 from Section~\ref{sec:Homotopy} dictate that to be a solution to \eqref{eq:Lasso_updated_homotopy}, $x^\star$ supported on $\Gamma$ with signs $z$ must obey
\begin{subequations}\label{eq:Lasso_update_opt}
\begin{gather}
A^T_{\Gamma}(A x^\star - y ) + \epsilon b^T_{\Gamma}(b x^\star - w) = -\tau z \label{eq:Lasso_update_opt1}\\
\|A^T_{\Gamma^c}(A x^\star - y ) + \epsilon b^T_{\Gamma^c}(b x^\star - w)\|_\infty <
\tau,  \label{eq:Lasso_update_opt2}
\end{gather}
\end{subequations}
Again, we can see the solution follows a piecewise linear path as $\epsilon$ varies, and the path changes directions at certain critical values of $\epsilon$ for which an element is either added or removed from the support of the solution.

Suppose we are at a solution $x_k$ to \eqref{eq:Lasso_updated_homotopy} at
one of these critical values of $\epsilon=\epsilon_k$.
Increasing $\epsilon$ an infinitesimal amount to $\epsilon_k^+$, we can subtract the optimality condition \eqref{eq:Lasso_update_opt1} at $x^\star=x_k$ from the condition for $x^\star=x_k^+$ to get
\begin{equation*}
\widetilde{ \partial x} = \begin{cases} -(\epsilon_{k}^+-\epsilon_k) (A^T_{\Gamma}A_\Gamma+ \epsilon_{k}^+ b^T_{\Gamma}b_\Gamma)^{-1} b^T_{\Gamma}(b x_k-w) & \text{on } \Gamma \\
 0 & \text{otherwise,} \end{cases}
\end{equation*}
where $\widetilde \delx = x_k^+-x_k$.

We can simplify this equation using the matrix inversion lemma, separating the step size from the update
direction.
Setting $U := A^T_\Gamma A_\Gamma+ \epsilon_k b_\Gamma^T
b_\Gamma$ and $u:= b_\Gamma U^{-1}b_\Gamma^T$, we have the following equations for the update direction
\begin{gather}
\delx = \begin{cases}  -U^{-1} b^T_\Gamma (b x_k -w) & \text{on } \Gamma \\
0 & \text{otherwise}\end{cases} \label{eq:Lasso_updated_delx}
\end{gather}
As $\epsilon$ increases from $\epsilon_k$, the solution moves in the direction $\partial x$.  However, unlike the update in Section~\ref{sec:DynamicX}, here the amount we move in the direction $\partial x$ is not proportional to the amount we change $\epsilon$; rather, moving from $\epsilon_k$ to $\epsilon_k^+$ will move the solution by $\theta_k\delx$, where
\[
\theta_k = \frac{\epsilon_{k}^+-\epsilon_k}{1+(\epsilon_{k}^+-{\epsilon_k})
u}.
\]

We now need to find the stepsize $\theta_k$ that will take us to the next critical point.  As we increase $\epsilon$ from $\epsilon_k$ (increasing $\theta_k$ from 0), the solution moves away from $x_k$ in direction $\delx$, until either an existing element in $x_k$ shrinks to zero or
one of the constraints in \eqref{eq:Lasso_update_opt2} becomes active.
The smallest step-size we can take such that an entry shrinks to zero is just
\begin{equation}
\label{eq:Lasso_updated_theta1}
\theta^- = \underset{j\in
\Gamma}{\text{min}} \left( \frac{-x_k(j)}{\delx  (j)} \right)_+,
\end{equation}
To find the smallest step size at which one of the inactive constraints becomes active, first note that as we move from $\epsilon_k$ to $\epsilon_k^+$, \eqref{eq:Lasso_update_opt} becomes
\begin{gather*}
\|A^T[A (x_k+\theta_k \delx) -y] + \epsilon_{k}^+ b^T[b (x_k+\theta_k \delx) - w]
\|_\infty \le \tau.
\end{gather*}
Setting
\begin{subequations}\label{eq:Lasso_updated_pk_dk}
\begin{gather}
p_k = A^T(A x_k-y) +\epsilon_k b^T (b x_k -w) \\
d_k = (A^T A+\epsilon_k b^T b)\partial x + b^T(b x_k-w),
\end{gather}
\end{subequations}
we are looking for the smallest $\theta_k$ such that $p_k(j) + \theta_k d_k(j) = \pm \tau$ for some $j\in\Gamma^c$.
This is given by
\begin{equation}
\label{eq:Lasso_updated_theta2}
\theta^+ = \underset{j\in \Gamma^c}{\text{min}} \left(
\frac{\tau-p_k(j)}{d_k(j)},
\frac{\tau+p_k(j)}{-d_k(j)}  \right)_+.
\end{equation}
The stepsize to the next critical point is then
\begin{equation}\label{eq:Lasso_updated_theta_choose}
\theta = \text{min} (\theta^+, \theta^-),
\end{equation}
and we set
\begin{equation}\label{eq:Lasso_updated_epsilon}
\epsilon_{k+1} = \epsilon_k + \frac{\theta}{1-\theta u},
\end{equation}
and $x_{k+1} = x_k + \theta \delx$. This procedure is repeated until $\epsilon=1$; pseudocode is given as Algorithm~\ref{alg:DynamicLasso} in Appendix~\ref{app:Pseudo-codes}.

We have to be a little cautious as we are tracking $\epsilon$ indirectly through the stepsize $\theta$.  In the last step of the algorithm, it is possible to choose $\theta$ large enough so that $\theta/(1-\theta u)$ is extremely large or negative.  In these situations, we simply reduce the value of $\theta$ until it corresponds to $\epsilon_{k+1}=1$, marking the endpoint of the solution path \cite{AR_L1filtering_Asilomar08}.

The main computational cost for each iteration of the algorithm is a rank-1 update for solving a $|\Gamma|\times|\Gamma|$ system of equations to find the direction $\delx$, and applications of $A$ and $A^T$ to find the stepsize.

\subsection{Dantzig selector update}
The homotopy formulation for \eqref{eq:DS_updated} is
\begin{equation}\label{eq:DS_updated_homotopy}
\underset{}{\text{minimize}} \;\; \|\tilde x\|_1 \;\; \text{subject to}
\;\; \|A^T(A\tilde x-y)+\epsilon b^T(b \tilde x-w)\|_\infty \le \tau,
\end{equation}
and the corresponding dual problem is
\begin{equation}\label{eq:DS_updated_homotopy_dual}
\text{maximize} \;\; -(\tau \|\lambda\|_1 + \langle \lambda, A^Ty+\epsilon
b^Tw\rangle) \quad \text{subject to} \quad \|A^TA\lambda + \epsilon b^T
b\lambda\|_\infty \le 1,
\end{equation}
where again varying $\epsilon$ from $0$ to $1$ takes us from the old solution to the new one.

The optimality conditions for  $(x^*, \lambda^*)$ to be a primal-dual solution pair
to \eqref{eq:DS_updated_homotopy} and \eqref{eq:DS_updated_homotopy_dual} at
some fixed value of $\epsilon$ and $\tau$ can be written as
\begin{subequations}\label{eq:DS_updated_opt_cond}
\begin{gather}
A^T_{\Gamma_\lambda}(A x^* -y) + \epsilon b^T_{\Gamma_\lambda}(b x^* -w) =
\tau
z_\lambda \label{eq:DS_opt_primal_supp}\\
A^T_{\Gamma_x}A \lambda^*+\epsilon b^T_{\Gamma_x}b \lambda^* = -z_x\label{eq:DS_opt_dual_supp}\\
\|A^T_{\Gamma_\lambda^c}(A x^* -y)  +\epsilon b^T_{\Gamma_\lambda^c}(b x^* -w) \|_\infty < \tau \label{eq:DS_opt_primal_off}\\
\|A^T_{\Gamma_x^c} A \lambda^*+ \epsilon b^T_{\Gamma_x^c}b \lambda^*
\|_\infty < 1,\label{eq:DS_opt_dual_off}
\end{gather}
\end{subequations}
where $\Gamma_x$ and $\Gamma_\lambda$ denote the supports of $x^*$ and
$\lambda^*$ respectively, and $z_x$ and $z_\lambda$ are the sign sequences on
their respective supports.

The procedure to trace the piecewise linear homotopy path is same as the BPDN update in principle, with the additional effort of keeping track of both the primal and dual variables at every homotopy step.
Assume that we have a solution $(x_k, \lambda_k)$ at some $\epsilon = \epsilon_k$ with support and sign sequence $(\Gamma_x, \Gamma_\lambda, z_x, z_\lambda)$.
As we increase $\epsilon$ away from $\epsilon_k$ to $\epsilon_k^+$, conditions
\eqref{eq:DS_opt_primal_supp} and \eqref{eq:DS_opt_dual_supp} tell us the primal and dual solutions will move according to
\begin{gather*}
\widetilde \delx = \begin{cases} -(\epsilon_k^+-\epsilon_k)(A^T_{\Gamma_\lambda}A_{\Gamma_x} + \epsilon_{k}^+ b^T_{\Gamma_\lambda}b_{\Gamma_x})^{-1} b^T_{\Gamma_\lambda}(b x_k-w) & \text{on } \Gamma_x \\
 0 & \text{otherwise} \end{cases},\\
\widetilde \dell= \begin{cases} -(\epsilon_k^+-\epsilon_k)(A^T_{\Gamma_x}A_{\Gamma_\lambda}+ \epsilon_{k}^+ b^T_{\Gamma_x}b_{\Gamma_\lambda})^{-1} b^T_{\Gamma_x}b \lambda_k & \text{on } \Gamma_\lambda \\
 0 & \text{otherwise} \end{cases}.
\end{gather*}
In the exact same manner, as with the BPDN update, the individual step sizes can be separated from the update directions using matrix inversion lemma.
We can write the solution values at $\epsilon_k^+$ as $x_k^+ = x_k + \theta_x \delx$ and $\lambda_k^+ = \lambda_k + \theta_\lambda \dell$, where $\theta_x$ and
$\theta_\lambda$ denote the step sizes and $\delx$ and $\dell$ the respective update directions. As we increase the step sizes $\theta_x$ and $\theta_\lambda$, $\epsilon$ increases and at some point there will be a change in {\em either} the primal support $\Gamma_x$ or the dual support $\Gamma_\lambda$. We pick the smallest step size, either $\theta_x$ or $\theta_\lambda$, which causes that change, and take primal and dual variables and constraints up to that point. This will give us the new critical value of $\epsilon$, the primal or dual solution at that critical point and the change in either $\Gamma_x$ or $\Gamma_\lambda$.
Depending on which variable, primal or dual, causes the change in support, we still have some room to change the other variable. So using the support update information we will update the other variable in a very similar way to the dual update in DS homotopy. For further details, see \cite{AR_PDpursuit_SPIE09, AR_l1_homotopy_webpage}. This procedure is also repeated until $\epsilon = 1$. 

\section{$\ell_1$ decoding}\label{sec:L1Decoding}

In this section, we will discuss a homotopy algorithm to update the solution
to the $\ell_1$ decoding problem \eqref{eq:L1_decode} as new measurements are added. %
We will use the language of a communications system: a transmitter is trying to send a message $x$ to a receiver.  The message is turned into a codeword by applying $A$, and the received signal $y=Ax+e$ is corrupted by a sparse error vector $e$.  The receiver recovers the message by solving \eqref{eq:L1_decode}.  If the codeword is long enough ($A$ has enough rows) and the error is sparse enough (not too many entries of $e$ are non-zero), the message will be recovered exactly.
The receiver will assume that the true message has been recovered when the error $Ax^*-y$ for the solution to \eqref{eq:L1_decode} has fewer than $m-n$ nonzero terms
(in general, the solution will contain exactly $m-n$ terms, and so this degeneracy indicates that the receiver has locked on to something special).
If the recovered error has exactly $m-n$ non-zero terms, the receiver asks the transmitter for more measurements (codeword elements).

Suppose that the receiver has just solved
\eqref{eq:L1_decode} to get a decoded message, and then $p$ new measurements of $x$ are received.
%
%
The updated system of equations is
\begin{equation}\label{eq:y=Ax+e_w=Bx+d_p}
\begin{bmatrix} y\\w \end{bmatrix} = \begin{bmatrix} A\\B \end{bmatrix} x +
\begin{bmatrix} e\\d \end{bmatrix},
\end{equation}
where $w$ represents $p$ new entries in the received codeword,
$B$ denotes $p$ new rows in the coding matrix, and $d$ is the error vector
for the new codeword entries. The receiver now must solve
the updated $\ell_1$ decoding problem
\begin{align}\label{eq:L1_decode_updated}
\underset{}{\text{minimize}} \; \|A\tilde x-y\|_1+\|B\tilde x-w\|_1.
\end{align}
These new measurements can be worked into the solution gradually, using the homotopy formulation
\begin{align}\label{eq:L1_decode_updated_homotopy}
\underset{}{\text{minimize}} \; \|A\tilde x-y\|_1+\epsilon \|B\tilde
x-w\|_1.
\end{align}

As in the Dantzig selector algorithms, we will find it convenient to trace the path of both the primal and dual solutions as $\epsilon$ increases from $0$ to $1$.
We begin by writing the dual of \eqref{eq:L1_decode_updated_homotopy} as
\begin{align}\label{eq:L1_decode_updated_homotopy_dual}
\underset{}{\text{maximize}}\;\; -\lambda^T y-\epsilon \nu^T w \quad
\text{subject to} \quad A^T \lambda + \epsilon B^T \nu= 0, \;\; \|\lambda\|_\infty \le 1, \; \|\nu \|_\infty \le 1,
\end{align}
where $\lambda\in \mathbb{R}^m$ and $\nu \in \mathbb{R}^p$ are the dual
optimization variables.

The optimality conditions for $(x_k, \lambda_k, \nu_k)$ to be a primal/dual solution set at $\epsilon = \epsilon_k$ can be derived as follows. Let
$e_k:=Ax_k - y$ and $d_k :=Bx_k -w$ be the error estimates for the first and second part of the codeword; denote their supports by $\Gamma_e$ and $\Gamma_d$ respectively.
Using the fact that the primal and dual objectives in \eqref{eq:L1_decode_updated_homotopy} and
\eqref{eq:L1_decode_updated_homotopy_dual} will be equal at their solutions,
we get the following conditions for $(x_k,\lambda_k, \nu_k)$:
\begin{subequations}\label{eq:L1_decode_opt}
\begin{align}
&\lambda_k = \text{sign}(Ax_k-y) \quad \text{on } \Gamma_e, \qquad
\|\lambda_k\|_\infty < 1 \quad \text{on }
\Gamma_e^c \label{eq:L1_decode_opt1}\\
&\nu_k = \text{sign} (Bx_k-w) \quad \text{on } \Gamma_d, \qquad  \|\nu_k\|_\infty <
1 \quad \text{on } \Gamma_d^c
\label{eq:L1_decode_opt2}\\
&A^T \lambda_k + \epsilon_k B^T \nu_k = 0. \label{eq:L1_decode_opt3}
\end{align}
\end{subequations}

The algorithm for tracking the solution to \eqref{eq:L1_decode_updated_homotopy},
\eqref{eq:L1_decode_updated_homotopy_dual} as $\epsilon$ moves from $0$ to $1$ consists of an initialization procedure followed by alternating updates of the primal and dual solution.  The critical points along the homotopy path correspond to the values of $\epsilon$ when an element enters or leaves the support of the estimate of the sparse error vector $[e^T~d^T]^T$.  We describe each of these stages below.

\paragraph*{{Initialization}}  We will use $x_0,~\lambda_0$ to denote the old primal and dual solutions at $\epsilon=0$; the old error estimate for the first $m$ codeword elements is $e_0:=Ax_0-y$.  We initialize the error estimate for the next $p$ elements as $d_0:=Bx_0-w$.  In general, if we have not yet recovered the underlying message, all of the terms in $d_0$ will be non-zero.  Throughout the algorithm, we will use $\Gamma$ as the index set for the error locations over all $m+p$ codeword elements; we initialize it with $\Gamma = \{\Gamma_e\cup\Gamma_d\}$, where $\Gamma_e$ is the support of $e_0$, and $\Gamma_d$ is the support of $d_0$.  The dual variable $\nu_0$ corresponding to these new measurements will start out as $\nu_0 = \operatorname{sign}(Bx_0-w)$.  Apart from keeping track of the support of the current error estimate, we will also find it necessary to keep track of which elements from the second part of the error $d$ have left the support at some time.  To this end, we initialize a set $\Gamma_n=\Gamma_d$, and when an element of $d$ shrinks to zero, we remove it from $\Gamma_n$ (we will never grow $\Gamma_n$).


Every step of the homotopy algorithm for $\ell_1$ decoding can be divided into two main parts: primal and dual update. Assume that we already have primal-dual solutions $(x_k,\lambda_k,\nu_k)$ for the problems in \eqref{eq:L1_decode_updated_homotopy} and
\eqref{eq:L1_decode_updated_homotopy_dual} at $\epsilon = \epsilon_k$, with
supports $\Gamma$ (corresponding to all non zero entries in the error
estimates) and $\Gamma_n$ (corresponding to entries of $d$ which remained
non-zero throughout the homotopy path so far). Let $e_k:=Ax_k-y$ and
$d_k:=Bx_k-w$ be the current error estimates.

\subsubsection{Dual update}
Assuming that the current error estimate has exactly $n$ terms which are zero (so $\Gamma$ has size $m+p-n$ and $\Gamma^c$ has size $n$), exactly $n$ entries in the dual vector $(\lambda_k,\nu_k)$ will have magnitude less than $1$.  Thus, there are $n$ degrees of freedom for which the dual solution can move during one step of the update; we will exercise this freedom by manipulating the dual coefficients on the set $\Gamma^c$.

If we combine both parts of the coding matrix together as
$G:=[A^T \; B^T]$ and both parts of the dual vector together as $\xi_k := [\lambda_k^T~~\nu_k^T]^T$, the optimality condition \eqref{eq:L1_decode_opt3} becomes
\begin{equation}\label{eq:L1_decode_feas_general}
G_{\Gamma_n^c} [\xi_k]_{\Gamma_n^c} + \epsilon_k G_{\Gamma_n}
[\xi_k]_{\Gamma_n} = 0.
\end{equation}
Increasing $\epsilon$ from $\epsilon_k$ to $\epsilon_k^+$, this condition for the new dual solution $\xi_k^+ = \xi_k + \widetilde {\partial \xi}$ can be written as
\begin{align}
G_{\Gamma_n^c} [\xi_k+\widetilde{\partial \xi}]_{\Gamma_n^c} + \epsilon_{k}^+ G_{\Gamma_n} [\xi_k+\widetilde{\partial \xi}]_{\Gamma_n} = 0\notag\\
G_{\Gamma_n^c} \widetilde{\partial
\xi}_{\Gamma_n^c}+(\epsilon_{k}^+-\epsilon_k) G_{\Gamma_n}
[\xi_k+\widetilde{\partial \xi}]_{\Gamma_n} = 0,
\label{eq:L1_decode_feas_dual_new}
\end{align}
where $\widetilde {\partial \xi}$ is supported only on the set $\Gamma^c$. Since $\Gamma_n \subset \Gamma$ and $\Gamma^c\subset \Gamma_n^c$, using \eqref{eq:L1_decode_feas_dual_new}, we can write the update direction $\partial
\xi$ and the step size $\theta_k^+$ required to
change $\epsilon$ from $\epsilon_k$ to $\epsilon_{k}^+$ as
\begin{equation}\label{eq:del_xi}
\partial \xi = \begin{cases} - (G_{\Gamma^c})^{-1} G_{\Gamma_n} [\xi_k]_{\Gamma_n} & \text{on } \Gamma^c \\
0 & \text{otherwise}, \end{cases}
\end{equation}
\begin{equation*}
\theta_k^+ = \epsilon_{k}^+- \epsilon_k.
\end{equation*}

As we increase $\epsilon$ from $\epsilon_k$, moving the solution in the direction $\partial \xi$, there will be a point at which an element of $\xi_k^+=\xi_k + \theta_k^+ \partial \xi$ will become active (equal to +1 or -1) on $\Gamma^c$. The smallest step size for this to happen can be computed as
\begin{gather}
\theta^+ = \underset{j \in \Gamma^c}{\text{min}} \left(
\frac{1-\xi_k(j)}{\partial \xi(j)}, \frac{1+\xi_k(j)}{-\partial \xi(j)}
\right)_+. \label{eq:L1_decode_theta_plus}
\end{gather}
The new values for $\epsilon$ and dual vector $\xi$ are given as
\begin{align*}
\epsilon_{k+1} = \epsilon_k + \theta^+, \qquad & \xi_{k+1} = \xi_k + \theta^+
\partial \xi.
\end{align*}
Let $\gamma^+$ be the index for the minimizer in \eqref{eq:L1_decode_theta_plus}.
This tells us that we have a new element in the estimated error vector at index $\gamma^+$ with sign $z_\gamma$, same as $\xi_{k+1}(\gamma^+)$.

\subsubsection{Primal update}
The dual update provides us with a new element in the support of the error estimate.  As the error estimate will have exactly $n$ entries which are zero until we have recovered the message, we know that one of elements currently in $\Gamma$ must shrink to zero.  This is accomplished by the primal update.

We have the following system of equations at $\epsilon =\epsilon_k$
\begin{equation}\label{eq:L1_decode_primal_ck}
\underbrace{\begin{bmatrix}A\\B \end{bmatrix}}_{G^T} x_k
-\underbrace{\begin{bmatrix}y\\w \end{bmatrix}}_s =
\underbrace{\begin{bmatrix}e_k\\d_k
\end{bmatrix}}_{c_k},
\end{equation}
where the old error estimate $c_k$ is supported only on the set $\Gamma$.
The dual update has indicated that our new error estimate will have a new active term at index $\gamma^+$, and that the sign of this new term will be $z_\gamma$.  Thus we need to update our estimate of the message $x$ such that the new error estimate $c_{k+1}$ has  $\operatorname{sign}[c_{k+1}(\gamma^+)]=z_\gamma$ and $c_{k+1}$ is zero at all
other indices in $\Gamma^c$.  In other words, an update direction $\widetilde \delx$ will satisfy
\begin{equation}\label{eq:L1_decode_feas_primal_new}
[G^T (x_k + \widetilde \delx) - s]_{\Gamma^c} = [c_k+\theta_k^- \partial c]_{\Gamma^c},
\end{equation}
where ${\partial c}$ is constrained on the set $\Gamma^c$ as
\begin{equation}\label{eq:L1_decode_delc}
{\partial c}\big|_{\Gamma^c} = \begin{cases} z_\gamma & \text{on } \gamma^+ \\
0 & \text{on } \Gamma^c \backslash \{\gamma^+\}
\end{cases}.
\end{equation}
We will choose $\theta^-_k$ above as the smallest value which shrinks an existing element in $c_k$ to zero; it will also be the unknown value for the new element in $c_{k+1}$ at index $\gamma^+$.

Using \eqref{eq:L1_decode_feas_primal_new} and \eqref{eq:L1_decode_delc} we can write the following system of equations to compute the update direction $\delx$
\begin{equation}\label{eq:L1_decode_delx}
[G^T]_{[\Gamma^c]} \delx = \begin{cases}z_\gamma & \text{on } \gamma^+\\
0 & \text{on } \Gamma^c \backslash\{\gamma^+\}\end{cases},
\end{equation}
where $[G^T]_{[\Gamma^c]}$ corresponds to the rows of $G^T=\begin{bmatrix}A \\
B\end{bmatrix}$ indexed by elements in the set $\Gamma^c$. We solve \eqref{eq:L1_decode_delx} to find $\delx$ and consequently $\partial c = G^T \delx$. The step size associated with  $\delx$ is $\theta^-_k$, and as we increase $\theta^-_k$ from 0, one of the elements in $c_k^+ = c_k + \theta_k^- \partial c$ will eventually shrink to zero. The value of this step size can be found with
\begin{gather}\label{eq:L1_decode_theta_minus}
\theta^- = \underset{j\in \Gamma}{\text{min}} \left( \frac{-c_k(j)}{{\partial
c}  (j)} \right)_+,
\end{gather}
which also gives the new value of $c_{k+1}(\gamma^+$). Let us denote $\gamma^-$ as the index corresponding to $\theta^-$. The new estimates for the message $x$ and error vector $c$ are given as
\begin{equation*}
x_{k+1} = x_k + \theta^- \delx \qquad c_{k+1} = c_k + \theta^-\partial c.
\end{equation*}

The support set can be updated as $\Gamma = [\Gamma \cup
\gamma^+]\backslash\{\gamma^-\}$. If at some point during primal update, an
element from within $\Gamma_n$ is removed, set $\Gamma_n = \Gamma_n \backslash
\{\gamma^-\}$ and $\xi_{k+1}(\gamma^-) = \epsilon_{k+1} \xi_{k+1}(\gamma^-)$.
Repeat this alternation of the dual and primal updates until $\epsilon$ becomes equal to 1.

The procedure outlined above used two working assumptions.  The first is that the error estimate will have exactly $n$ zero entries until we recover the original message $x$.  The second is that any $n\times n$ submatrix formed by picking $n$ rows from the $m+p\times n$ coding matrix will be nonsingular.  The second assumption allows us to calculate the update directions for both the primal and dual; the first ensures that this update direction is unique.
Both of these assumptions are true with probability $1$ if the coding matrix is Gaussian or a random projection, and they are true with very high probability if the coding matrix $A$ is Bernoulli \cite{Malioutov_2008_CSsequential}.
In addition to this, the condition number of these submatrices will be \emph{fairly} controlled \cite{RudelsonVershynin_2008_Littlewood}.
The algorithm can be extended to properly handle situations where these assumptions do not hold, but we will not discuss this here.

As before, the main computational cost in this algorithm comes from one matrix-vector product to compute $\partial c$ and rank-1 update for solution of a $|\Gamma| \times |\Gamma|$ system to find the update directions $\partial \xi$ and $\delx$.




\section{Robust $\ell_1$ decoding}\label{sec:RobustL1Decoding}

In practice, we would like a decoding scheme that can handle codewords which have been corrupted both by a small number of gross errors and a small amount of ambient noise.  In \cite{CandesRandall_2008_RobustEC}, an optimization program similar to \eqref{eq:Robust_EC} (or \eqref{eq:Robust_EC_Lasso}) was proposed for accomplishing this type of robust error correction.
In this section we will discuss the updating procedure for these problems as new elements of the codeword are received.

Assume that we have solved \eqref{eq:Robust_EC_Lasso} for the system in \eqref{eq:y=Ax+e_RobustDecode} and then we receive $p$ new measurements: $w =Bx+d+q_w$, where $B$ denotes $p$ new rows in the coding matrix, $d$ denotes the sparse errors and $q_w$ denotes small noise. The updated system is
\begin{equation}\label{eq:y=Ax+e_RobustDecode_update}
\begin{bmatrix}y\\w \end{bmatrix} = \begin{bmatrix}A\\B
\end{bmatrix}x+\begin{bmatrix} e \\
d\end{bmatrix}+\begin{bmatrix}q_y\\q_w
\end{bmatrix},
\end{equation}
the new decoding program becomes
\begin{align}\label{eq:Robust_EC_updated}
{\text{minimize}} \quad \tau (\|\tilde {e}\|_1 + \|\tilde{d}\|_1)+
\frac{1}{2}(\|\tilde{q}_y\|^2_2+ \|\tilde q_w\|_2^2)\quad  \text{subject
to} \quad & A\tilde x+\tilde{e}+\tilde{q}_y = y  \\
& B\tilde x + \tilde d + \tilde q_w = w.\notag
\end{align}
The homotopy formulation (with parameter $\epsilon$) to work in the new measurement is
\begin{align}\label{eq:Robust_EC_homotopy}
{\text{minimize}} \;\; \tau (\|\tilde {e}\|_1 + \epsilon \|\tilde{d}\|_1)+
\frac{1}{2}(\|\tilde{q}_y\|^2_2+ \|\tilde q_w\|_2^2)\quad \text{subject to}
\quad & A\tilde
x+\tilde{e}+\tilde{q}_y = y \\
& B\tilde x + \tilde d + \tilde q_w = w.\notag
\end{align}
Similar to \eqref{eq:Robust_EC_Lasso} we can form
a BPDN type equivalent problem to \eqref{eq:Robust_EC_homotopy}:
\begin{equation}\label{eq:Robust_EC_Lasso_homotopy}
\underset{}{\text{minimize}} \;\;\tau (\|\tilde
e\|_1+\epsilon \|\tilde d\|_1) +
\frac{1}{2}\left\|P\left(\begin{bmatrix}\tilde e\\\tilde
d\end{bmatrix}-\begin{bmatrix}y\\w\end{bmatrix}\right)\right\|_2^2,
\end{equation}
where $P$ is the matrix whose rows span the null space of $F^T$, i.e., $PF = 0$, where $F:=\begin{bmatrix}A\\B\end{bmatrix}$.

Note that while the decoding problem \eqref{eq:Robust_EC_Lasso_homotopy} has the same form as the BPDN, the homotopy formulation \eqref{eq:Robust_EC_homotopy} is significantly different than those in Sections~\ref{sec:DynamicX} and \ref{sec:Dynamic_seq}.
The difference is due to the fact that here the size of the sparse entity we wish to estimate (the error) grows with the number of measurements.


In order to build the homotopy path, we need the optimality conditions for the
solution to \eqref{eq:Robust_EC_Lasso_homotopy}. The necessary condition
for a pair $(e_k,d_k)$ to be a solution to \eqref{eq:Robust_EC_Lasso_homotopy} at
$\epsilon=\epsilon_k$ is
\begin{equation*}
\left|P^TP\left(\begin{bmatrix}e_k\\d_k\end{bmatrix}-\begin{bmatrix}y\\w\end{bmatrix}\right)\right|
\preceq
\begin{bmatrix}\tau\\\epsilon_k \tau\end{bmatrix},
\end{equation*}
where $\preceq$ denotes the componentwise inequality; the last inequalities, involving $\epsilon_k$, correspond to the non-zero elements in $d_k$. We collect both parts of the error estimate together as $c_k:=[e_k^T ~ d_k^T]^T$ and both parts of the measurements as $s:=[y^T~w^T]^T$.
The support of $c_k$ is given as $\Gamma:=[\Gamma_e\cup \Gamma_n]$, where $\Gamma_n$ is the index set corresponding to those elements of $d_k$ which remain non-zero in $c_k$ and $\Gamma_e$ is the index set for the remaining non-zero entries in $c_k$. Let $z_e$ and $z_d$ be the sign sequence of $c_k$ on $\Gamma_e$ and $\Gamma_n$ respectively. The optimality conditions can now be written as
\begin{subequations}\label{eq:Robust_EC_opt}
\begin{gather}
P_{\Gamma_e}^TP(c_k -s) = -\tau z_e \label{eq:Robust_EC_opt1}\\
P_{\Gamma_n}^TP(c_k -s) = -\epsilon_k\tau z_d \label{eq:Robust_EC_opt2}\\
\|P_{\Gamma^c}^TP(c_k -s) \|_\infty < \tau. \label{eq:Robust_EC_opt3}
\end{gather}
\end{subequations}

We find the update direction by examining these optimality conditions as we increase
$\epsilon$ a small ways from $\epsilon_k$.
The solution $c_k^+$ at $\epsilon = \epsilon_k^+$ must obey
\begin{gather*}
P_{\Gamma}^TP(c_k^+ -s) =
\begin{bmatrix} -\tau z_e \\ -\epsilon_{k}^+ \tau
z_d\end{bmatrix},
\end{gather*}
and so
\begin{gather*}
P_{\Gamma}^TP (c_k^+-c_k) = \begin{bmatrix} 0 \\ -(\epsilon_{k}^+-\epsilon_k) \tau
z_d\end{bmatrix}.
\end{gather*}
Since $c_k^+$ and $c_k$ are both supported on the set $\Gamma$, we can write the update direction $\partial c = c_k^+-c_k$ and associated step size $\theta_k$ which moves $\epsilon$ from $\epsilon_k$ to $\epsilon_k^+$ as
\begin{gather}
\partial c = \begin{cases} -
(P_{\Gamma}^TP_{\Gamma})^{-1}\begin{bmatrix}0 \\
z_d\end{bmatrix} & \text{on } \Gamma \\
0 & \text{otherwise} \end{cases}\label{eq:Robust_EC_delc}\\
\theta_k = (\epsilon_{k}^+-\epsilon_k)\tau. \notag 
\end{gather}

Finally, we need to find the stepsize $\theta$ that will take us to the next critical value of $\epsilon$. As we increase $\epsilon$ from $\epsilon_k$, the solution $c_k$ moves in the direction $\partial c$ until either an element in $c_k$ shrinks to zero or one of the constraints in \eqref{eq:Robust_EC_opt3} become actives (equal to $\tau$). The smallest amount we can move $\epsilon$ so that an element in $c_k$ shrinks to zero is
\begin{equation}\label{eq:Robust_EC_theta1}
\theta^- = \underset{j\in \Gamma}{\text{min}} \left( \frac{-c_k(j)}{{\partial
c}  (j)}
\right)_+.
\end{equation}
For the smallest step size that activates a constraint, set
\begin{subequations}\label{eq:Robust_EC_pk_dk}
\begin{gather}
p_k = P^TP(c_k-s)\\
d_k = P^TP \partial c,
\end{gather}
\end{subequations}
and find the smallest $\theta^+$ so that $p_k(j) + \theta^+ d_k(j) = \pm \tau$ for some $j\in \Gamma^c$.  In other words,
\begin{equation}\label{eq:Robust_EC_theta2}
\theta^+ = \underset{j\in \Gamma^c}{\text{min}} \left(
\frac{\tau-p_k(j)}{d_k(j)},
\frac{\tau+p_k(j)}{-d_k(j)}\right)_+.
\end{equation}
The stepsize to the next critical point is then
\begin{equation}\label{eq:Robust_EC_theta_choose}
\theta = \text{min}(\theta^+,\theta^-).
\end{equation}

With the direction $\partial c$ and stepsize $\theta$ calculated, the next critical value of $\epsilon$ is
\begin{equation*}
\epsilon_{k+1} = \epsilon_k + \frac{\theta}{\tau},
\end{equation*}
and the solution (error estimate) at $\epsilon_{k+1}$ is
\begin{equation*}
c_{k+1} = c_k + \theta \partial c,
\end{equation*}
with one element either entering or leaving the support.

Repeat this procedure until $\epsilon$ becomes equal to 1. If at any point an element of $d_k$ from $\Gamma_n$ shrinks to zero, we remove it from $\Gamma_n$ and treat it as if it were an element of $e_k$ (i.e., without homotopy). If all the elements in $\Gamma_n$ shrink to zero, we will be able to quit.
Pseudocode for this procedure is given as Algorithm~\ref{alg:Robust_EC} in Appendix~\ref{app:Pseudo-codes}. The final solution
$\widehat{c}$ can be used to find the decoded message $\widehat{x}$ using
\begin{equation*}
\widehat x = (F^TF)^{-1}F^T(s-\widehat c).
\end{equation*}

The main computational cost involves computing the kernel matrix $P$ in the start and solve \eqref{eq:Robust_EC_delc} for $\partial c$ at each homotopy step. Computing matrix $P$ will cost $O(mn^2)$ for the first step, and afterwards with each new measurement computing any such matrix $P$ will take only a few rank one updates. Since only one element changes in $\Gamma$ at every homotopy step, the update direction $\partial c$ can also be computed efficiently using few rank one update.

Our discussion above assumes the invertibility of $P_{\Gamma}^TP_{\Gamma}$.  Recall that $P^T$ is the matrix whose columns span the left null space of $F$, (e.g.,
$P = I - F(F^TF)^{-1}F^T$). For $P_{\Gamma}^TP_{\Gamma}$ to be singular requires that a vector with sparsity strictly less than $m-n+p$ be in the null space of $P$.  This will not be true for generic coding matrices $F$: if we chose $F$ to be a random projection or iid Gaussian matrix, $P_{\Gamma}^T P_{\Gamma}$ will be invertible for all $\Gamma$ with $|\Gamma| \leq m-n+p$ with probability one.

%

\section{Numerical examples}\label{sec:Numerical_Examples}

In this section we will discuss some simulation results which demonstrate the
efficiency of our proposed dynamic update.  A MATLAB implementation of each of the algorithms discussed in the paper, along with the experiments presented below, is available online at \cite{AR_l1_homotopy_webpage}.

\subsection{Time varying sparse signals}

We will first look at the update algorithm presented in Section~\ref{sec:DynamicX} for reconstructing a series of sparse signals.  The algorithm is most effective when the support of the solution does not change too much from instance to instance.

In the examples below, we start with a sparse signal $x\in \mathbb{R}^n$ and
its $m$ measurements according to the model in \eqref{eq:y=Ax+e}. We first
solve \eqref{eq:Lasso} for a given value of $\tau$. Then the signal is perturbed slightly to $\breve x$ , a new set of
$m$ measurements $\breve y = A\breve x+\breve e$ are taken, and  \eqref{eq:DynamicX_Lasso} is solved using
Algorithm~\ref{alg:DynamicX}. In all of the examples below, we have used an $m\times n$ Gaussian matrix as our measurement matrix $A$, with all entries
independently distributed $\mathrm{Normal}(0,1/m)$.

To gauge how the difference in support will effect the speed of the update, we start with a synthetic example.  In this first simulation, we start with a sparse signal $x$ which contains $\pm 1$ spikes at randomly chosen $K$ locations.  The measurement vector $y$ is generated as in \eqref{eq:y=Ax+e}, with
$e$ as a Gaussian noise whose entries are distributed
$\mathrm{Normal}(0,0.01^2)$.  We solve \eqref{eq:Lasso} for a given value of
$\tau$.
Then we modify the sparse signal $x$ to get $\breve x$ as follows.  First, we perturb the non-zero entries of $x$ by adding random numbers distributed
$\mathrm{Normal}(0,0.1^2)$.  Then $K_n$ new entries are added to $x$, with the locations chosen uniformly at random, and the values distributed $\mathrm{Normal}(0,1)$.  New measurements $\breve{y}:= A\breve x + \breve e$ are generated, with another realization of the noise vector $\breve e$, and \eqref{eq:DynamicX_Lasso} is solved using the DynamicX algorithm (Algorithm~\ref{alg:DynamicX}).

The results of 500 simulations with $n=1024,~ m = 512,~ K = m/5$ are summarized in Table~\ref{table:DynamicX}.  In each simulation, $K_n$ was selected uniformly from  $[0, K/20]$.  Several values of $\tau$ were tested, $\tau =
\lambda\|A^Ty\|_\infty$ with $\lambda \in \{0.5, ~0.1, ~0.05, ~0.01\}$.
The experiments were run on a standard desktop PC, and two numbers were recorded: the average number of times we needed to apply\footnote{
Each iteration of the DynamicX algorithm requires an application of $A^TA$ along with several much smaller matrix-vector multiplies to perform the rank-1 update.  Since these smaller matrix-vector multiplies are so much cheaper, the numbers in the table include only applications of the full $A^TA$.}
$A^T$ and $A$ (nProdAtA), and the average CPU time needed to complete the experiment (CPU).

Table~\ref{table:DynamicX} also compares DynamicX to three other methods.  The first is ``Standard BPDN homotopy'', which resolves \eqref{eq:DynamicX_Lasso} from scratch using our own implementation of the homotopy algorithm reviewed in Section~\ref{sec:Homotopy} (starting $\tau$ large and gradually reducing it to its desired value).  The second is the GPSR-BB algorithm \cite{Figueiredo_2007_GPSR},
%
%
which is ``warm started'' by using the previously recovered signal as the starting point.  The third algorithm is FPC\_AS
\cite{FPC_AS}, which is also warm started.  The accuracy in GPSR and FPC was chosen so that the relative error between the exact solution and their solution was $10^{-6}$.  We see that DynamicX compares favorably across a large range of $\tau$.

A few comments about Table~\ref{table:DynamicX} are in order.  First, the DynamicX solves \eqref{eq:DynamicX_Lasso} to within machine precision, while both GPSR and FPC are iterative algorithms providing approximate solutions; we accounted for this fact by having a rather stringent accuracy requirement.  This level of accuracy is important for signals which have high dynamic range (some elements of $x$ are much bigger than others).  However, there are many situations in which less accurate solutions will suffice, and the number of matrix products required for GPSR and FPC will be reduced.  Second, we feel that the number of applications of $A^TA$ is a more telling number than the CPU time, as the latter can be affected significantly by the implementation.



\begin{figure}[t]
    \centering
  \includegraphics[width=1\columnwidth, angle=0]{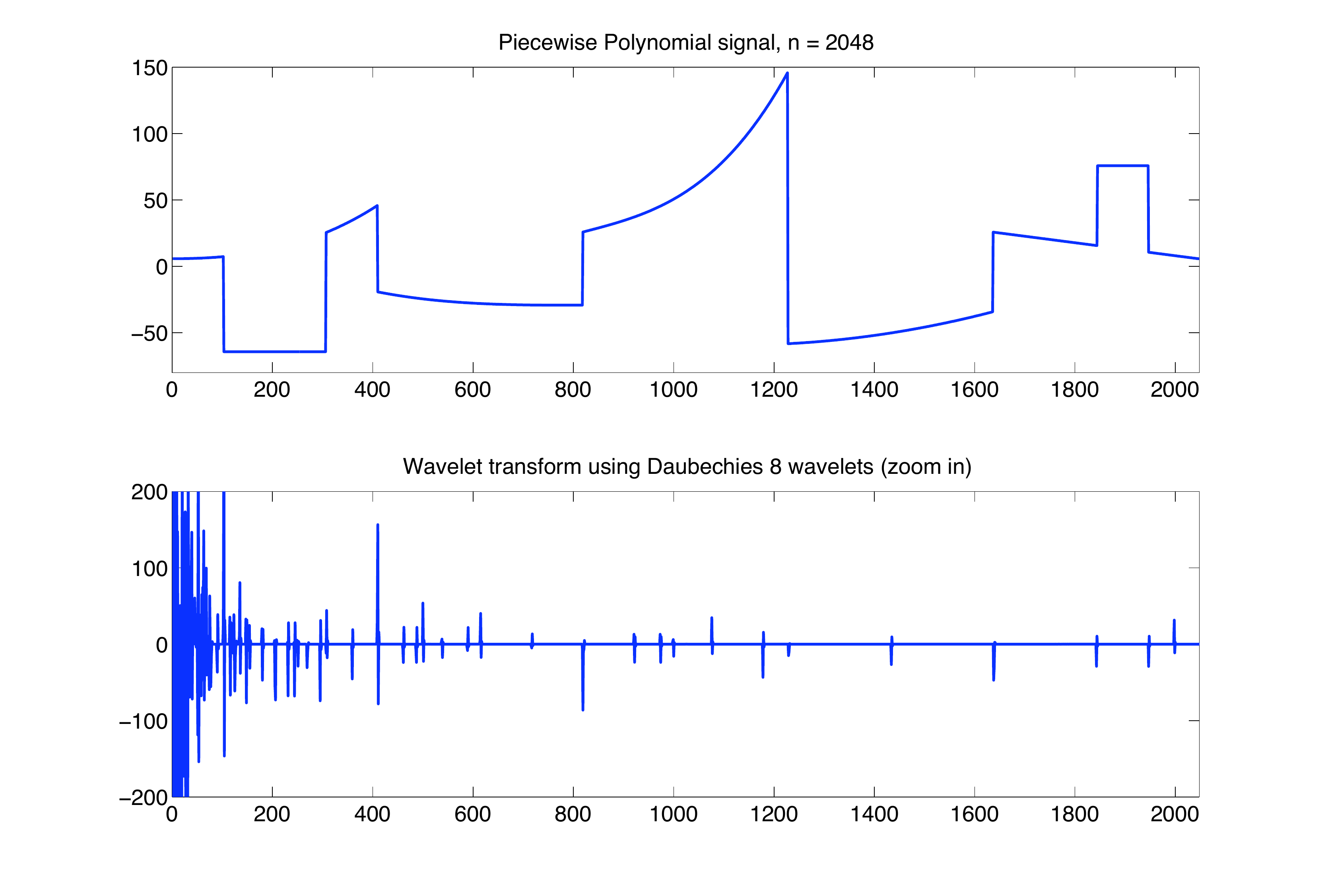}\\
  \caption{An example of Piecewise smooth signal, sparse in wavelet domain}
  \label{fig:PcwPoly}
\end{figure}

\begin{figure}[t]
    \centering
  \includegraphics[width=1\columnwidth, angle=0]{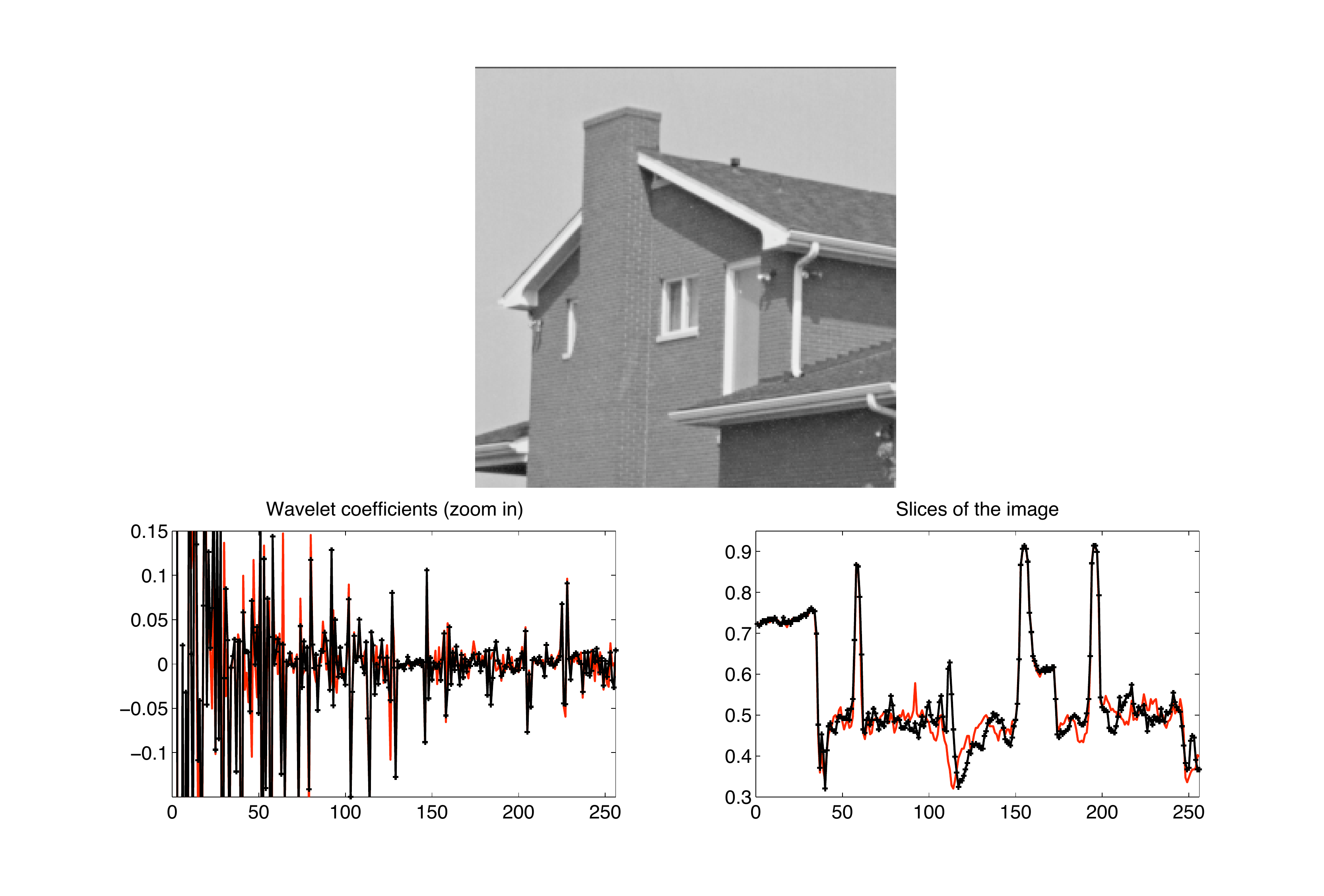}\\
  \caption{An image of house (256x256): lower images represent two consecutive slices from this image (on the right) and their respective wavelet transform coefficients (on the left)}
  \label{fig:House}
\end{figure}

Table~\ref{table:DynamicX} also contains results for three other experiments with the following descriptions.

\noindent{\bf Blocks:} In this experiment, we recover a series of 200 piecewise constant signals of length $n=2048$, similar to the {\em Blocks} signal from WaveLab \cite{WaveLab}.  We use the Haar wavelet transform to represent the signal, and take $m=1024$ measurements.  Each signal is a slight variation of the last: the discontinuities stay fixed, while the levels of the constant regions are perturbed by multiplying by a random number uniformly distributed between $0.8$ and $1.2$.  As the signal varies, the signs and locations of the significant wavelet coefficients vary as well.

\noindent{\bf Piecewise polynomial:}  This experiment is similar to the Blocks experiment, except that we use a piecewise polynomial (cubic) signal and represent it using the Daubechies 8 wavelet transform.  A typical signal and its wavelet transform are shown in Figure~\ref{fig:PcwPoly}.  The polynomial functions are perturbed from signal to signal by adding small Gaussian random variables to the polynomial coefficients.

\noindent{\bf Slices of the {\em House} image:} In this experiment, we take the 256 column slices of the House image, shown in Figure~\ref{fig:House}, as our sequence of signals, and use the Haar wavelet transform to represent them.  As the singularities will move slightly from slice to slice, more of the support in the wavelet domain will change, making this a more challenging data set than the previous examples.

\begin{table*}[]
\caption{Comparison of the dynamic update of time-varying sparse signals  using
the standard BPDN homotopy, GPSR and FPC.  Results are given in terms of the number of products with $A^T$ and $A$, and CPU time.}
\begin{center}\label{table:DynamicX}
\begin{tabular}{|l|c|c|c|c|c|}
\hline
\multirow{2}{*} {Signal type} & $\lambda$&  DynamicX & Standard Homotopy & GPSR-BB  & FPC\_AS\\
& $(\tau = \lambda \|A^Ty\|_\infty)$ & (nProdAtA, CPU)& (nProdAtA, CPU)& (nProdAtA, CPU)& (nProdAtA, CPU)\\
\hline
$n=1024,$ & 0.5  & (11.84, 0.031) & (42.05, 0.10) & (15.34, 0.03) & (31.29, 0.055) \\ \cline{2-6}
$m=512,$& 0.1 & (12.9, 0.055) & (154.5, 0.491)  & (54.45, 0.095) & (103.38, 0.13) \\
\cline{2-6}
$K=m/5,$  & 0.05 &  (14.56, 0.062) & (162, 0.517) & (58.17, 0.10) & (102.37, 0.14) \\ \cline{2-6}
 values: $\pm 1$ spikes   & 0.01 &  (23.72, 0.132) & (235, 0.924) & (104.5, 0.18) & (148.65, 0.177) \\ \cline{2-6}
 \hline
\text{Blocks} & 0.01 & (2.7,0.028)& (76.8,0.490)& (17,0.133) & (53.5,0.196)
\\\cline{2-6} \hline
\text{Pcw. Poly.} & 0.01 & (13.83,0.151)& (150.2,1.096)&
(26.05, 0.212) & (66.89, 0.250) \\\cline{2-6} \hline
\text{House slices} & 0.005 &   (44.69, 0.022)& (76.85,0.03)& (220.49, 0.03) & (148.96, 0.055)\\
\hline
\end{tabular}
\end{center}
\end{table*}


\subsection{Sequential measurements}
In this experiment our underlying signal $x$ contains $\pm 1$ spikes at $K$ randomly chosen locations. The $m\times n$ measurement matrix $A$ is Gaussian with entries distributed $\mathrm{Normal}(0,1/m)$.  We observe $y=Ax+e$ with the entries of $e$ iid Gaussian with zero mean and variance $10^{-4}$. We start by solving \eqref{eq:Lasso} for a given value of $\tau$. We add one new measurement $w=bx+d$, where $b$ is a row vector whose entries are distributed
as those in $A$ and $d$ is the additional noise term, and update the solution using the DynamicSeq algorithm (Algorithm~\ref{alg:DynamicLasso}).  The results are summarized in Table~\ref{table:DynamicLasso}, and are compared as before against the standard BPDN homotopy algorithm, GPSR with a warm start, and FPC with a warm start.

The average number of homotopy iterations taken for the update varies
with the sparsity of the solution. At large values of $\tau$, the
solution has a small number of non-zero entries and the update requires something like 2 or 3 homotopy steps.
For smaller values of $\tau$, the solution has many more
non-zero terms and the number of iterations in the update increases; for example, at $\tau = 0.01 \|A^Ty\|_\infty$
an average 8 homotopy steps were required to incorporate a new measurement.

\begin{table*}[]
\caption{Comparison of Dynamic BPDN update, GPSR and FPC with one new measurement}
\begin{center}\label{table:DynamicLasso}
\begin{tabular}{|l|c|c|c|c|c|}
\hline
\multirow{2}{*} {Signal type} & $\lambda$&  DynamicSeq & Standard Homotopy & GPSR-BB  & FPC\_AS\\
 & $(\tau = \lambda \|A^Ty\|_\infty)$ & (nProdAtA, CPU)& (nProdAtA, CPU)& (nProdAtA, CPU)& (nProdAtA, CPU)\\
\hline
  $n=1024,$& 0.5 &  (2.43, 0.007) & (42.1, 0.10) & (12.21, 0.02) & (23.84, 0.032) \\ \cline{2-6}
   $m=512$& 0.1 &  (4.27, 0.019) & (151.6, 0.491)  & (40.28, 0.07) & (104.84, 0.11) \\ \cline{2-6}
     $K=m/5, $ & 0.05 &  (5.57, 0.024) & (161.6, 0.537) & (42.3, 0.072) & (119.2, 0.12) \\ \cline{2-6}
    values: $\pm 1$ spikes&0.01 &  (8.3, 0.05) & (231, 0.929) & (56.6, 0.095) & (141.4, 0.145) \\\cline{2-6}
 \hline
\end{tabular}
\end{center}
\end{table*}

\subsection{Robust $\ell_1$ decoding}
Now we will look at an example for the robust error correction update algorithm from Section~\ref{sec:RobustL1Decoding}. We start with an
arbitrary signal $x\in \mathbb{R}^n$ with $n=150$; we generate $x$ by drawing its entries from a standard normal distribution.
The initial coding matrix $A$ is generated by drawing
an $m\times n$ Gaussian matrix and orthogonalizing the columns, where $m=300$.
The sparse error $e$ is added to the codeword $Ax$
by selecting $K=60$ random locations in $Ax$ and setting those values to zero. The small noise $q_y$ is added to all locations of the codeword; its entries are distributed $\mathrm{Normal}(0,0.01^2)$.
The program \eqref{eq:Robust_EC_Lasso} is solved for $\tau=0.01$ with $Q = I-A(A^TA)^{-1}A^T$, giving us an initial solution.
We add $p$ new elements to the corrupted codeword, deciding whether or not to corrupt any new observation (set it to zero) by drawing an independent Bernoulli random variable that has a 10\% probability of success.  The solution is then updated using Algorithm~\ref{alg:Robust_EC}.

\begin{table}[!h]
\caption{Average number of homotopy steps and CPU time taken to update the robust $\ell_1$ decoding solution.  The ``cold start'' columns calculate the new solution from scratch using the standard BPDN homotopy algorithm, while the ``warm start'' columns use Algorithm~\ref{alg:Robust_EC} to update the solution.}
\begin{center}\label{table:REC}
\begin{tabular}{|c|c|c|c|c|}
\hline
\multirow{2}{*}{New entries } & \multicolumn{2}{|c|}{Time per iteration (in sec.)} & \multicolumn{2}{|c|}{Homotopy steps per iteration} \\
\cline{2-5}
$(p)$& cold start & warm start & cold start & warm start \\
\hline 1 & 0.275   & 0.041 &    180.33  & 16.44   \\\hline 2  &   0.325 &
0.078  & 182.27 &  26.29  \\\hline 5  &0.292 &  0.109 & 175.27 &  40.05  \\
\hline 10 &0.255  &  0.144 & 176.15  & 58.64 \\\hline
\end{tabular}
\end{center}
\end{table}

Table~\ref{table:REC} compares the average number of homotopy steps and CPU time for the update for {$p \in \{1, 2, 5, 10\}$}.
Note that the average number of steps scales favorably with $p$: adding $10$ measurements at once requires $58.64$ iterations to update the solution (an average of $5.86$ per entry), while adding $1$ measurement at a time requires $16.44$ iterations on average.
Likewise, the average time per entry when $p=10$ is $0.144/10 = 0.0144$ seconds, as compared to $0.041$ for $p=1$.  These numbers suggest that it is advantageous to add the measurements in blocks rather than one at a time.

\section{Conclusions}
We have presented a suite of homotopy algorithms to quickly update the solution to a variety of $\ell_1$ minimization programs.  The updates can occur when either new measurements are added to the system or the signal we are observing changes slightly.
The homotopy methods discussed are simple and inexpensive, and promise significantly lower marginal cost than re-solving an entirely new optimization program.
These methods break the update down into a series of linear steps.
The computational cost of each step is a few matrix-vector multiplications, and simulation results show that for reasonably sparse signals, only a small number of steps are required for the update.
These algorithms are extremely efficient in cases where support of the solution does not change much. The numerical results further show that for dynamic update, homotopy methods are superior to \emph{warm started} GPSR and FPC methods.

%

%


\bibliographystyle{ieeetr}
\bibliography{CS-bib}

\newcommand{\noopsort}[1]{} \newcommand{\printfirst}[2]{#1}
  \newcommand{\singleletter}[1]{#1} \newcommand{\switchargs}[2]{#2#1}
\begin{thebibliography}{10}

\bibitem{CandesTao_2006_NearOptimal}
E.~J. Cand\`es and T.~Tao, ``Near-optimal signal recovery from random
  projections: Universal encoding strategies?,'' {\em Information Theory, IEEE
  Transactions on}, vol.~52, no.~12, pp.~5406--5425, Dec. 2006.

\bibitem{Candes_2006_StableRecovery}
E.~Cand\`es, J.~Romberg, and T.~Tao, ``{Stable signal recovery from incomplete
  and inaccurate measurements},'' {\em Comm. Pure Appl. Math}, vol.~59, no.~8,
  pp.~1207--1223, 2006.

\bibitem{Tropp_2006_JustRelax}
J.~Tropp, ``{Just relax: Convex programming methods for identifying sparse
  signals in noise},'' {\em Information Theory, IEEE Transactions on}, vol.~52,
  no.~3, pp.~1030--1051, 2006.

\bibitem{Donoho_2006_StableRecoveryOvercomplete}
D.~Donoho, M.~Elad, and V.~Temlyakov, ``{Stable recovery of sparse overcomplete
  representations in the presence of noise},'' {\em Information Theory, IEEE
  Transactions on}, vol.~52, no.~1, pp.~6--18, 2006.

\bibitem{Donoho_2006_CS}
D.~Donoho, ``Compressed sensing,'' {\em Information Theory, IEEE Transactions
  on}, vol.~52, no.~4, pp.~1289--1306, April 2006.

\bibitem{CandesPlan_2008_NearIdealModelSelection}
E.~Cand\`es and Y.~Plan, ``{Near-ideal model selection by $\ell_1$
  minimization},'' {\em Annals of Statistics (to appear)}, 2008.

\bibitem{Zhu_2008_StableRecoveryIT}
C.~Zhu, ``Stable recovery of sparse signals via regularized minimization,''
  {\em Information Theory, IEEE Transactions on}, vol.~54, pp.~3364--3367, July
  2008.

\bibitem{CandesTao_2007_DS}
E.~Cand\`es and T.~Tao, ``{The Dantzig selector: Statistical estimation when
  $p$ is much larger than $n$},'' {\em Annals of Statistics}, vol.~35, no.~6,
  pp.~2313--2351, 2007.

\bibitem{CandesTao_2005_DecodingLP}
E.~Cand\`es and T.~Tao, ``Decoding by linear programming,'' {\em Information
  Theory, IEEE Transactions on}, vol.~51, no.~12, pp.~4203--4215, Dec. 2005.

\bibitem{rudelson05ge}
M.~Rudelson and R.~Vershynin, ``Geometric approach to error correcting codes
  and reconstruction of signals,'' {\em International Mathematics Research
  Notices}, no.~64, pp.~4019--4041, 2005.

\bibitem{Golub_1996_MatrixComputation}
G.~Golub and C.~Van~Loan, {\em {Matrix Computations}}.
\newblock Johns Hopkins University Press, 1996.

\bibitem{Bjorck_1996_NumericalLS_book}
{\AA}.~Bj{\"o}rck, {\em {Numerical Methods for Least Squares Problems}}.
\newblock Society for Industrial Mathematics, 1996.

\bibitem{Chen_99_BasisPursuit}
S.~S. Chen, D.~L. Donoho, and M.~A. Saunders, ``Atomic decomposition by basis
  pursuit,'' {\em SIAM Journal on Scientific Computing}, vol.~20, no.~1,
  pp.~33--61, 1999.

\bibitem{Tibshirani_1996_LASSO}
R.~Tibshirani, ``{Regression shrinkage and selection via the lasso},'' {\em
  Journal of the Royal Statistical Society, Series B}, vol.~58, no.~1,
  pp.~267--288, 1996.

\bibitem{KimBoyd_2007_L1_ls}
S.-J. Kim, K.~Koh, M.~Lustig, S.~Boyd, and D.~Gorinevsky, ``An interior-point
  method for large-scale $\ell_1$-regularized least squares,'' {\em Selected
  Topics in Signal Processing, IEEE Journal of}, vol.~1, no.~4, pp.~606--617,
  2007.

\bibitem{Figueiredo_2007_GPSR}
M.~Figueiredo, R.~Nowak, and S.~Wright, ``Gradient projection for sparse
  reconstruction: Application to compressed sensing and other inverse
  problems,'' {\em Selected Topics in Signal Processing, IEEE Journal of},
  vol.~1, no.~4, pp.~586--597, 2007.

\bibitem{hale2008fpc}
E.~Hale, W.~Yin, and Y.~Zhang, ``{Fixed-Point Continuation for
  $\ell_1$-minimization: Methodology and Convergence},'' {\em SIAM Journal on
  Optimization}, vol.~19, p.~1107, 2008.

\bibitem{Yin_2008_Bregman}
W.~Yin, S.~Osher, D.~Goldfarb, and J.~Darbon, ``Bregman iterative algorithms
  for $\ell_1$ minimization with application to compressed sensing,'' {\em SIAM
  Journal on Imaging sciences}, vol.~1, no.~1, pp.~143--168, 2008.

\bibitem{Efron_2004_LARS}
B.~Efron, T.~Hastie, I.~Johnstone, and R.~Tibshirani, ``{Least angle
  regression},'' {\em Annals of Statistics}, vol.~32, no.~2, pp.~407--499,
  2004.

\bibitem{OsbornePresnell_2000_NewApproachLasso}
M.~Osborne, B.~Presnell, and B.~Turlach, ``{A new approach to variable
  selection in least squares problems},'' {\em IMA Journal of Numerical
  Analysis}, vol.~20, no.~3, pp.~389--403, 2000.

\bibitem{l1magic}
E.~Cand\`es and J.~Romberg, ``{$\ell_1$-\textsc{magic}: Recovery of Sparse
  Signals via Convex Programming}.'' {http://www.acm.caltech.edu/l1magic/}.

\bibitem{James_2007_DASSO}
G.~James, P.~Radchenko, and J.~Lv, ``{The DASSO algorithm for fitting the
  Dantzig selector and the Lasso},'' {\em {Journal of the Royal Statistical
  Society, Series B}}, vol.~71, pp.~127--142, 2009.

\bibitem{Asif_2008_MSThesis}
M.~S. Asif, ``{Primal Dual Pursuit: A homotopy based algorithm for the Dantzig
  selector},'' Master's thesis, {Georgia Institute of Technology}, August 2008.

\bibitem{CandesRandall_2008_RobustEC}
E.~J. Cand\`es and P.~A. Randall, ``Highly robust error correction by convex
  programming,'' {\em Information Theory, IEEE Transactions on}, vol.~54,
  no.~7, pp.~2829--2840, 2008.

\bibitem{Garrigues_2008_NIPS}
P.~J. Garrigues and L.~E. Ghaoui, ``{An homotopy algorithm for the Lasso with
  online observations},'' {\em Neural Information Processing Systems (NIPS)
  21}, December 2008.

\bibitem{AR_l1_homotopy_webpage}
M.~S. Asif and J.~Romberg, ``{$\ell_1$ Homotopy : A MATLAB toolbox for homotopy
  algorithms in $\ell_1$ norm minimization problems}.''
  http://users.ece.gatech.edu/$\sim$sasif/homotopy.

\bibitem{Malioutov_2005_HomotopyContinuation}
D.~Malioutov, M.~Cetin, and A.~Willsky, ``Homotopy continuation for sparse
  signal representation,'' {\em {IEEE International Conference on Acoustics,
  Speech, and Signal Processing,}}, vol.~5, pp.~v/733--v/736, March 2005.

\bibitem{Bertsekas_1999_NonlinearProg_book}
D.~Bertsekas, {\em {Nonlinear programming}}.
\newblock Athena Scientific Belmont, Mass, 1999.

\bibitem{Fuchs_2004_OnSparseRep}
J.~Fuchs, ``{On sparse representations in arbitrary redundant bases},'' {\em
  Information Theory, IEEE Transactions on}, vol.~50, no.~6, pp.~1341--1344,
  2004.

\bibitem{Boyd_book_ConvexOptimization}
S.~Boyd and L.~Vandenberghe, {\em {Convex Optimization}}.
\newblock {Cambridge University Press}, March 2004.

\bibitem{lustig2007sma}
M.~Lustig, D.~Donoho, and J.~Pauly, ``{Sparse MRI: The application of
  compressed sensing for rapid MR imaging},'' {\em Magnetic Resonance in
  Medicine}, vol.~58, no.~6, pp.~1182--1195, 2007.

\bibitem{CotterRao_2002_SparseChannelEst}
S.~F. Cotter and B.~D. Rao, ``{Sparse channel estimation via matching pursuit
  with application to equalization},'' {\em Communications, IEEE Transactions
  on}, vol.~50, no.~3, pp.~374--377, 2002.

\bibitem{AR_L1filtering_Asilomar08}
M.~S. Asif and J.~Romberg, ``{Streaming measurements in compressive sensing:
  $\ell_1$ filtering},'' {\em 42nd Asilomar conference on Signals, Systems and
  Computers}, October 2008.

\bibitem{AR_PDpursuit_SPIE09}
M.~S. Asif and J.~Romberg, ``Dantzig selector homotopy with dynamic
  measurements,'' {\em Proc. IS\&T/ SPIE Computational Imaging VII}, vol.~7246,
  no.~1, p.~72460E, 2009.

\bibitem{Malioutov_2008_CSsequential}
D.~Malioutov, S.~Sanghavi, and A.~Willsky, ``Compressed sensing with sequential
  observations,'' {\em {IEEE International Conference on Acoustics, Speech, and
  Signal Processing,}}, pp.~3357--3360, April 2008.

\bibitem{RudelsonVershynin_2008_Littlewood}
M.~Rudelson and R.~Vershynin, ``{The Littlewood--Offord problem and
  invertibility of random matrices},'' {\em Advances in Mathematics}, 2008.

\bibitem{FPC_AS}
Z.~Wen and W.~Yin, ``{FPC\_{}AS: A MATLAB Solver for $\ell_1$-Regularized Least
  Squares Problems}.''
  {http://www.caam.rice.edu/$\sim$optimization/L1/FPC\_{}AS/}.

\bibitem{WaveLab}
J.~Buckheit, S.~Chen, D.~Donoho, and I.~Johnstone, ``{Wavelab 850, Software
  toolbox}.'' {http://www-stat.stanford.edu/$\sim$wavelab/}.

\end{thebibliography}

\newpage
\appendices
\section{Pseudo-codes}\label{app:Pseudo-codes}
\algsetup{indent = 2em}
\begin{algorithm}[]
\caption{Dynamic update of time varying sparse signal: DynamicX\_BPDN}\label{alg:DynamicX}
\begin{algorithmic}
\STATE Start with $\epsilon_0 = 0$ at solution $x_0$ to \eqref{eq:Lasso} with
support $\Gamma$ and sign sequence $z$ on the $\Gamma$ for $k=0$.
\REPEAT
 \STATE compute $\delx$
 as in
 \eqref{eq:DynamicX_delx} \STATE compute $p_k, d_k$ as in
 \eqref{eq:DynamicX_pk_dk}
 and $\theta$ as in \eqref{eq:dynx_mintheta}
 \STATE
 $x_{k+1} = x_k
 + \theta \delx $
  \STATE $\epsilon_{k+1} = \epsilon_k + \theta$
 \IF{$\epsilon_{k+1} > 1$}
 \STATE $\theta = 1-\epsilon_k$
 \STATE $x_{k+1} = x_k + \theta \delx$
 \STATE $\epsilon_{k+1} = 1$
 \STATE \texttt{break;} \hfill \COMMENT{Quit without any further update}
 \ENDIF
 \IF{$\theta = \theta^-$}
 \STATE $\Gamma \leftarrow \Gamma~ \backslash ~\{\gamma^-\}$
  \ELSE
 \STATE $\Gamma \leftarrow \Gamma \cup \{\gamma^+\}$
 \ENDIF
\STATE $ k \leftarrow k+1$
\UNTIL{\texttt{stopping criterion is satisfied}}
\end{algorithmic}
\end{algorithm}

\algsetup{indent = 2em}
\begin{algorithm}[p]
\caption{Dynamic update with sequential measurements: DynamicSeq\_BPDN}\label{alg:DynamicLasso}
\begin{algorithmic}
\STATE Start with $\epsilon_0 = 0$ at solution $x_0$ to \eqref{eq:Lasso} with
support $\Gamma$ and sign sequence $z$ on the $\Gamma$ for $k=0$.
\REPEAT
 \STATE compute $\delx$
 as in
 \eqref{eq:Lasso_updated_delx} \STATE compute $p_k, d_k$ as in \eqref{eq:Lasso_updated_pk_dk}
 and $\theta$ as in \eqref{eq:Lasso_updated_theta_choose}
 \STATE
 $x_{k+1} = x_k
 + \theta \delx $
  \STATE $\epsilon_{k+1} = \epsilon_k + \dfrac{\theta}{1-\theta u}$
 \IF{$\epsilon_{k+1} > 1$} 
 \STATE $\theta =
 \dfrac{1-\epsilon_k}{1+(1-\epsilon_k)u}$
 \STATE $x_{k+1} = x_k + \theta \delx$
 \STATE $\epsilon_{k+1} = 1$
 \STATE \texttt{break;} \hfill \COMMENT{Quit without any further update}
 \ENDIF
 \IF{$\theta = \theta^-$}
 \STATE $\Gamma \leftarrow \Gamma~ \backslash ~\{\gamma^-\}$
  \ELSE
 \STATE $\Gamma \leftarrow \Gamma \cup \{\gamma^+\}$
 \ENDIF
\STATE $ k \leftarrow k+1$
\UNTIL{\texttt{stopping criterion is satisfied}}
\end{algorithmic}
\end{algorithm}

\algsetup{indent = 2em}
\begin{algorithm}[p]
\caption{$\ell_1$ Decoding Homotopy}\label{alg:L1Decoding}
\begin{algorithmic}
\STATE Start at $\epsilon_0 = 0$ with primal-dual solution $x_0$, $\lambda_0$,
error estimate $e_0:=Ax_0-y$ with support $\Gamma_e$. Set $\Gamma_n$ as the set
of indices corresponding to the $p$ new measurements, set $d_0:=Bx_0-w$, and
$\nu_0:=z_d$. Set $\Gamma = [\Gamma_e \cup \Gamma_n]$, $c_0:=
\begin{bmatrix}e_0\\ d_0\end{bmatrix}$, $\xi_0:=
\begin{bmatrix}\lambda_0\\\nu_0\end{bmatrix}$ and
$G:=[A^T \; B^T]$.
\REPEAT
\STATE \textbf{Dual update:} \STATE compute $\partial \xi$ as in
\eqref{eq:del_xi}
 \STATE find $\theta^+$, $\gamma^+$ and $z_\gamma$ as described in \eqref{eq:L1_decode_theta_plus}
 \STATE $\xi_{k+1} = \xi_k + \theta^+ \partial \xi $
 \STATE $\epsilon_{k+1} = \epsilon_k + \theta^+$
 \IF{$\epsilon_{k+1} > 1$}
 \STATE $\theta^+ = 1-\epsilon_k$
 \STATE $\xi_{k+1} = \xi_k + \theta^+ \partial \xi$
 \STATE $\epsilon_{k+1} = 1$
 \STATE \texttt{break;} \hfill \COMMENT{Quit without any further update}
 \ENDIF
\STATE \textbf{Primal update:}
\STATE compute $\delx$ from \eqref{eq:L1_decode_delx}, set $\partial c := G^T\delx$ \STATE find $\theta^-$ and $\gamma^-$ as described in \eqref{eq:L1_decode_theta_minus}
\STATE $x_{k+1} = x_k + \theta^-\delx$
\STATE $c_{k+1} = c_k + \theta^-\partial c$
\STATE $\Gamma \leftarrow [\Gamma \cup \gamma^+]\backslash \{\gamma^-\}$ \IF{$\gamma^- \in \Gamma_n$}
\STATE $\Gamma_n \leftarrow \Gamma_n \backslash \{\gamma^-\}$ \hfill\COMMENT{Treat the corresponding error location without homotopy}
\STATE $\xi_{k+1}(\gamma^-) = \epsilon_{k+1} \xi_{k+1}(\gamma^-)$
\IF{$\Gamma_n$ becomes empty}
\STATE \texttt{break;} \hfill \COMMENT{Lucky breakdown}
\ENDIF \ENDIF \STATE $ k \leftarrow k+1$
\UNTIL{\texttt{stopping criterion is satisfied}}
\end{algorithmic}
\end{algorithm}

\algsetup{indent = 2em}
\begin{algorithm}[p]
\caption{Robust $\ell_1$ decoding Homotopy}\label{alg:Robust_EC}
\begin{algorithmic}
\STATE Start at $\epsilon_0 = 0$ with solution $(x_0,e_0)$ to
\eqref{eq:Robust_EC_Lasso}. Define $d_0:=w-Bx_0$, $c_0 :=\begin{bmatrix}e_0\\
d_0\end{bmatrix}$ with support $\Gamma:=[\Gamma_e \cup \Gamma_n]$, where
$\Gamma_e$ and $\Gamma_n$ are the supports of $e_0$ and $d_0$ respectively. Let $z_e$ be sign of $e_0$ on $\Gamma_e$ and $z_d$ be sign of $d_0$. Define $F:=\begin{bmatrix}A\\ B\end{bmatrix}$ and
compute $P$.
\REPEAT
 \STATE compute $\partial c$ as in
 \eqref{eq:Robust_EC_delc}
 \STATE compute $p_k, d_k$ as in \eqref{eq:Robust_EC_pk_dk}
 and $\theta$ as in \eqref{eq:Robust_EC_theta_choose}
 \STATE
 $c_{k+1} = c_k
 + \theta \partial c $
  \STATE $\epsilon_{k+1} = \epsilon_k + \dfrac{\theta}{\tau}$
 \IF{$\epsilon_{k+1}\ge 1$}
 \STATE $\theta =
{1-\epsilon_k}\tau$
 \STATE $c_{k+1} = c_k + \theta \partial c$
 \STATE $\epsilon_{k+1} = 1$
 \STATE \texttt{break;} \hfill \COMMENT{Quit without any further update}
 \ENDIF
 \IF{$\theta = \theta^-$}
 \STATE $\Gamma \leftarrow \Gamma~ \backslash ~\{\gamma^-\}$
 \IF{$\gamma^- \in \Gamma_n$}
\STATE $\Gamma_n \leftarrow \Gamma_n \backslash \{\gamma^-\}$ \hfill\COMMENT{Treat the corresponding error location without homotopy}
\IF{$\Gamma_n$ becomes empty}
\STATE \texttt{break;} \hfill \COMMENT{Lucky breakdown}
 \ENDIF
 \ENDIF
  \ELSE
 \STATE $\Gamma \leftarrow \Gamma \cup \{\gamma^+\}$
\ENDIF
\STATE $ k \leftarrow k+1$
\UNTIL{\texttt{stopping criterion is satisfied}}
\STATE $\widehat x = (F^TF)^{-1}F^T(s-c_{k+1})$ \hfill \COMMENT{Decoded
dataword}
\end{algorithmic}
\end{algorithm} 

\end{document}